\documentclass[12pt]{article}

\makeatletter
\renewcommand{\@cite}[2]{{#1\if@tempswa , #2\fi}}
\renewcommand{\@biblabel}[1]{$^{#1\hfill }\!\!$}
\makeatother

\newcommand{\citeup}[1]{${\!}^{\cite{#1}}$}

\newcommand{\sth}{\sigma^3}
\newcommand{\stw}{\sigma^2}
\newcommand{\son}{\sigma^1}

\newcommand{\pmS}{\,^{\pm}\!S}
\newcommand{\pS}{\,^{+}\!S}
\newcommand{\mS}{\,^{-}\!S}
\newcommand{\pmbvstw}{\,^{\pm}\!\mbox{\boldmath$v$}_{\sigma^2}}
\newcommand{\pmRstw}{\,^{\pm}\!R_{\sigma^2}}
\newcommand{\pbvstw}{\,^{+}\!\mbox{\boldmath$v$}_{\sigma^2}}
\newcommand{\pRstw}{\,^{+}\!R_{\sigma^2}}
\newcommand{\mbvstw}{\,^{-}\!\mbox{\boldmath$v$}_{\sigma^2}}
\newcommand{\mRstw}{\,^{-}\!R_{\sigma^2}}
\newcommand{\Rstw}{R_{\sigma^2}}
\newcommand{\Omsth}{\Omega_{\sigma^3}}
\newcommand{\sumstw}{\sum_{\sigma^2}}
\newcommand{\prodsth}{\prod_{\sigma^3}}
\renewcommand{\d}{{\rm d}}
\renewcommand{\Re}{{\mbox{Re\,}}}
\renewcommand{\Im}{{\mbox{Im\,}}}
\newcommand{\D}{{\cal D}}

\newcommand{\N}{{\cal N}}
\newcommand{\tilN}{\tilde{\cal N}}
\newcommand{\Ngade}{{\cal N}_{\gamma\dot{\delta}}}
\newcommand{\Malbegade}{{\cal M}^{\alpha\dot{\beta}}_{\gamma\dot{\delta}}}
\newcommand{\Ialbegade}{{\cal I}^{\alpha\dot{\beta}}_{\gamma\dot{\delta}}}
\newcommand{\tilIalbegade}{\tilde{\cal I}^{\alpha\dot{\beta}}_{\gamma\dot{\delta}}}
\newcommand{\tilNgade}{\tilde{\cal N}_{\gamma\dot{\delta}}}
\newcommand{\tilIalga}{\tilde{I}^{\alpha}_{\gamma}}
\newcommand{\tilIdotbedotde}{\tilde{I}^{\dot{\beta}}_{\dot{\delta}}}

\newcommand{\pmbphi}{\,^{\pm}\!\mbox{\boldmath$\phi$}}

\newcommand{\pbv}{\,^{+}\!\mbox{\boldmath$v$}}
\newcommand{\mbv}{\,^{-}\!\mbox{\boldmath$v$}}
\newcommand{\bv}{\mbox{\boldmath$v$}}
\newcommand{\bvstw}{\mbox{\boldmath$v$}_{\sigma^2}}
\newcommand{\bu}{\mbox{\boldmath$u$}}
\newcommand{\bw}{\mbox{\boldmath$w$}}
\newcommand{\rw}{\mbox{\rm w}}
\newcommand{\rv}{\mbox{\rm v}}
\newcommand{\rsv}{\mbox{\scriptsize\rm v}}
\newcommand{\bn}{\mbox{\boldmath$n$}}
\newcommand{\rbn}{\mbox{\rm\bf n}}
\newcommand{\bsn}{\mbox{\scriptsize\boldmath$n$}}

\newcommand{\br}{\mbox{\boldmath$r$}}
\newcommand{\bsr}{\mbox{\scriptsize\boldmath$r$}}
\newcommand{\bs}{\mbox{\boldmath$s$}}
\newcommand{\bq}{\mbox{\boldmath$q$}}
\newcommand{\bvarphi}{\mbox{\boldmath$\varphi$}}
\newcommand{\bpsi}{\mbox{\boldmath$\psi$}}
\newcommand{\bphi}{\mbox{\boldmath$\phi$}}

\newcommand{\pmvstw}{\,^{\pm}\!v_{\sigma^2}}

\newcommand{\vstw}{v_{\sigma^2}}
\newcommand{\vaonajbonbk}{v^{a_1}...v^{a_j}(v^{b_1}...v^{b_k})^*}
\newcommand{\naonajbonbk}{n^{a_1}...n^{a_j}(n^{b_1}...n^{b_k})^*}
\newcommand{\rconclamdondmu}{r_{c_1}...r_{c_\lambda}(r_{d_1}...r_{d_\mu})^*}
\newcommand{\vaonajvjbonbkvk}{{v^{a_1}...v^{a_j} \over \mbox{\rm v}^j}{(v^{b_1}...v^{b_k})^* \over \mbox{\rm v}^{*k}}}
\newcommand{\pmOm}{\,^{\pm}\!\Omega}
\newcommand{\pmR}{\,^{\pm}\!R}
\newcommand{\pR}{\,^{+}\!R}

\newcommand{\dfuns}{$\delta$-functions }
\newcommand{\desi}{\delta^6}
\newcommand{\deth}{\delta^3}
\newcommand{\detw}{\delta^2}
\newcommand{\de}{\delta}
\newcommand{\beon}{\mbox{\boldmath$e$}_1}
\newcommand{\betw}{\mbox{\boldmath$e$}_2}
\newcommand{\sh}{{\rm sh}}
\newcommand{\ch}{{\rm ch}}
\newcommand{\ddh}{{{\rm d}\over {\rm d}h}}
\newcommand{\ddhstar}{{{\rm d}\over {\rm d}h^*}}
\newcommand{\ddz}{{{\rm d}\over {\rm d}z}}
\newcommand{\ddzstar}{{{\rm d}\over {\rm d}z^*}}
\newcommand{\twi}{{2\over i}}
\newcommand{\varphinp}{\varphi^{(n,p)}}
\newcommand{\hz}{h(z) \propto z}
\newcommand{\hzarcz}{h(z) \propto z$ or $\arcsin z}
\newcommand{\harcz}{h(z) \propto \arcsin z}

\begin{document}

\title{Attributing sense to some integrals in Regge calculus}
\author{V.M. Khatsymovsky \\
 {\em Budker Institute of Nuclear Physics} \\ {\em
 11 Lavrentyev ave,
 Novosibirsk,
 630090,
 Russia}
\\ {\em e-mail: khatsym@inp.nsk.su}}
\date{}
\maketitle
\begin{abstract}
Regge calculus minisuperspace action in the connection representation has the form in which each term is linear over some field variable (scale of area-type variable with sign). We are interested in the result of performing integration over connections in the path integral (now usual multiple integral) as function of area tensors even in larger region considered as independent variables. To find this function (or distribution), we compute its moments, i. e. integrals with monomials over area tensors. Calculation proceeds through intermediate appearance of $\delta$-functions and integrating them out. Up to a singular part with support on some discrete set of physically unattainable points, the function of interest has finite moments. This function in physical region should therefore exponentially decay at large areas and it really does being restored from moments. This gives for gravity a way of defining such nonabsolutely convergent integral as path integral.
\end{abstract}


\newpage

\noindent {\large\bf 1. Introduction}

\noindent Strict definition of the functional integral is possible for Gaussian case; for small deviations from this case it is considered to be definable perturbatively. For general relativity system perturbative expansion is poorly defined due to nonrenormalizability of gravity, and we have nonGaussian path integral. The action is essentially nonlinear, but in the Cartan-Weyl form, in terms of tetrad and connections, the action can be viewed as linear in some field variable which is bilinear in the tetrad. This issuing feature of gravity inherent in some form also in minisuperspace formulations is important for what follows.

Since we do not possess exact definition of nonGaussian functional integral, we need its finite dimensional realization on minisuperspace system. Piecewise flat manifold or simplicial complex provides such framework known as Regge calculus \citeup{Regge,Cheeger}. Invoking the notion of discrete tetrad and connection first considered in Ref. \cite{Fro} we have suggested in Ref. \cite{Kha} representation of the minisuperspace Regge action in terms of area tensors and finite rotation SO(4) (SO(3,1) in the Minkowsky case) matrices, and also in terms of (anti-)selfdual parts of finite rotation matrices. For the latter we write
\begin{equation}
\pmS = \sumstw \sqrt{\pmbvstw^2} \arcsin {\pmbvstw * \pmRstw (\Omega )\over \sqrt{\pmbvstw^2} }.
\end{equation}

\noindent Here $\pmbvstw$ are vectors parameterizing (anti-)selfdual parts $\pmvstw^{ab}$ of the bivector $\vstw^{ab}$ of the triangle $\stw$ ($v^{ab} = {1\over 2}\epsilon^{ab}_{~~cd}l^c_1l^d_2$ for some two 4-vectors $l^c_1$, $l^d_2$ which span the triangle), $\sqrt{\pmbvstw^2}$ is area of the triangle, in the Minkowsky case $\Omsth$ is rotation SO(3,1) matrix on the tetrahedron $\sth$ which we call simply connection, $\Rstw$ is curvature matrix on the triangle $\stw$ (holonomy of $\Omega$'s). For a 3-vector $\bv$ and a $3\times 3$ matrix $R$ we have denoted $\bv * R \equiv {1\over 2}v^a R^{bc} \epsilon_{abc}$, and for $\pmRstw$, the (anti-)selfdual part of $\Rstw$, we have used adjoint, SO(3) representation (to be precise, SO(3,C) matrix).

The sense of the considered representations is that upon excluding rotation matrices by classical equations of motion these result in the same Regge action (that is, on-shell). Taking into account that in the Minkowsky case $\pS = ( \mS )^*$ we can write out the most general combination of $\pS$, $\mS$ which i) reduces to Regge action on-shell and ii) is real, as $S$ = $C \pS + C^* \mS$ where $C + C^* = 1$, that is $C = 1/2 + i \cdot \mbox{(real parameter)}$. At the same time, in the continuum theory the {\it Holst} action which generalizes the Cartan-Weyl form of the Einstein action \citeup{Holst,Fat} is easily seen to have the form $(1 + i/\gamma) \pS_{\rm cont} + (1 - i/\gamma) \mS_{\rm cont}$ where $\pmS_{\rm cont}$ are (anti-)selfdual parts of the Cartan-Weyl continuum action, $\gamma$ is known as Barbero-Immirzi parameter \citeup{Barb,Imm}. Therefore we can write $C = (1 + i / \gamma)/2$ where the discrete analog of $\gamma$ is denoted by the same letter. We assume $0 < \gamma < \infty$.

Consider such the action $S$ and discretized functional integral $\int \exp (iS) D q$, $q$ are field variables (some factors of the type of Jacobians could also be present). Functional integral approach in Regge calculus was earlier developed, see, e. g., Refs. \cite{Fro,HamWil1,HamWil2}. Suppose we have performed integration over rotation matrices and are interested in the dependence of the intermediate result on area tensors. Of course, different area tensors are not independent, but nothing prevent us from studying analytical properties in the extended region of varying these area tensors as if these were independent variables. Namely, consider integral
\begin{eqnarray}\label{intDOm}                                                     %1
\N = \int \exp {i\over 2} \sumstw \left [ \left (1 + {i\over \gamma} \right ) \sqrt{\pbvstw^2} \arcsin {\pbvstw * \pRstw (\Omega )\over \sqrt{\pbvstw^2} } \right. \nonumber\\ \left. + \left (1 - {i\over \gamma} \right ) \sqrt{\mbvstw^2} \arcsin {\mbvstw * \mRstw (\Omega )\over \sqrt{\mbvstw^2} } \right ] \prodsth \D \Omsth.
\end{eqnarray}

\noindent Matrices $\pmOm$, $\pmR$ can be parameterized by complex vector angles $\pmbphi = \bvarphi \mp i\bpsi$ (rotation by the angle $\sqrt{\pmbphi^2}$ around the unit vector $\pmbphi /\sqrt{\pmbphi^2}$).

We regard (\ref{intDOm}) as function of arbitrary $\pbv$, $\mbv = (\pbv)^*$ which we redenote as $\bv, \bv^*$ in the main body of the paper. To be specific, we study the following integrals,
\begin{equation}\label{N-gamma-delta}                                              %2
\Ngade (\bv,\bv^*) \equiv \int \exp \left [{i\over 2}\rv h(\bn\br) + {i\over 2}\rv^*h(\bn\br)^*\right ] \rconclamdondmu \D R.
\end{equation}

\noindent Here $\gamma = (c_1...c_\lambda)$, $\dot{\delta} = (\dot{d}_1 ... \dot{d}_\mu)$ are multiindices; the dot on an index has the only sense that corresponding vector component enters complex conjugated. The $h(z)$ is analytical at $z=0$ odd function $h(z) = -h(-z)$. Principal value $\arcsin z$ or simply $z$ are examples of $h(z)$. Besides that, $\rv = \sqrt{\bv^2}$, $\bn = \bv / \rv$, $\bn^2 = 1$, $r_a = \epsilon_{abc}\pR^{bc}/2 = \phi_a(\sin \phi)/\phi$, $\phi = \sqrt{\bphi^2}$, $\bphi = \bvarphi - i\bpsi$,
\begin{equation}\label{DR}                                                         %3
\D R = \left ({1\over \sqrt{1-\br^2}} - 1\right )\left ({1\over \sqrt{1-\br^{*2}}} - 1\right ) {\d^3\br \d^3\br^* \over (8\pi^2)^2 \br^2 \br^{*2}}.
\end{equation}

\noindent Here $\d^3\br \d^3\br^* \equiv 2^3 \d^3 \Re \br \d^3 \Im \br \equiv 2^3 \d^6 \br$. The monomial $\rconclamdondmu$ originates as a term in Taylor expansion of possible dependence on $R$ of the factors provided by $\Rstw$ in other triangles due to the Bianchi identities. As usual, functional integral is not absolutely convergent. The additional complications could be connected with growth (exponential) of the Haar measure on Lorentz boosts. Then the result of integration over the latter might be defined as a generalized function, or distribution, rather than an ordinary function. For example, the following integral diverges but could be defined as distribution,
\begin{equation}\label{distrib}
\int \exp (ivr) r^n \d r = 2 \pi (-i)^n \delta^{(n)} (v),
\end{equation}

\noindent namely, as Fourier transform of $r^n$ treated as another distribution. Then it is appropriate to study instead of equation (\ref{distrib}) the result of integrating both parts of it with suitable probe functions. For the latter we chose those ones for which corresponding integrals could be easily defined.
Let us issue from the integral of $\Ngade$ with powers of $\bv, \bv^*$,
\begin{equation}\label{Malbegade}                                                  %4
\Malbegade (l,m) = \int \Ngade (\bv, \bv^*)(\bv^2)^l (\bv^{*2})^m \vaonajbonbk \d^6 \bv,
\end{equation}

\noindent and change overall integration order: first integrate over $\d^6 \bv$, then over $\d^6 \br$. Here $\d^6 \bv \equiv \d^3 \Re \bv \d^3 \Im \bv$, etc. The $\alpha, \dot{\beta}$ are multiindices. The only sense of distinguishing between superscripts and subscripts is that the former refer to $\bv, \bv^*$, the latter refer to $\br, \br^*$. Call (\ref{Malbegade}) the moment of $\Ngade$ (specified by $\alpha, \beta, l, m$).

Note that the case $h(z) \propto z$ can be of significance as well as $\arcsin z$, and probably even more related to canonical approach to constructing the functional integral measure. Namely, we could ask whether form of the full discrete path integral (i. e. simply many-fold integral) exists which results in the canonical integral form $\int \exp (iS(p,q)) \prod_t \d p (t) \d q (t)$ in the continuous time limit when we shrink the edges along any direction chosen as time $t$ and pass to the canonical formalism with conjugate pairs $p, q$. More generally, gravity action is of the form $S(p,q,\lambda)$ with non-dynamical $\lambda$. Then standard path integral derivation gives the form $\int \exp (iS(p,q,\lambda)) \!$ $\! \prod_t \!$ $\! \d p (t) \!$ $\! \d q (t) \!$ $\! \d \lambda (t)$ on condition that $S$ is linear in $\lambda$, $S = \int (p \dot{q} - \sum_{\alpha} \lambda_{\alpha} \Phi_{\alpha} (p,q)) \d t$, and $\Phi_{\alpha} (p,q)$ mutually commute w.r.t. Poisson brackets. The latter just takes place in the 3 dimensional case, in accordance with Waelbroeck's derivation of the commuting constraints in general discrete 3-dimensional gravity system \citeup{Wael}. This allows us to define the discrete path integral form of interest \citeup{Kha2}. A particular point in this derivation is that both genuine Regge action $S$ and that one $\tilde{S}$ differing from $S$ by omitting the 'arcsin' functions (i. e. by replacing $h(z) \propto \arcsin z \to z$) are equivalent on-shell due to the local triviality of the 3 dimensional gravity. Of these namely $\tilde{S}$ in the continuous time limit has the above form $S(p,q,\lambda)$ linear in $\lambda$ and therefore just appears in the exponential.

In the reminder of the present paper we define the moments of the integrals over connections of interest, show that support for the singular distributional part of these integrals restored from moments lays outside the physical region $\Im \rv^2 = 0$, separate out the regular part and present it for the simplest integral.

\medskip

\noindent {\bf\large 2. Defining moments of the path integral distribution}

\noindent At $h(z) \propto z$ when calculating $\Malbegade (l, m)$ (\ref{Malbegade}) we get derivatives of \dfuns $\desi (\br )\!$ $\!\equiv\!$ $\!\deth (\Re \br)\!$ $\!\cdot\!$ $\!\deth (\Im \br)\!$ which are then integrated over $\D R$. Finiteness is provided by analyticity of this measure at $\br, \br^* \to 0$ w.r.t. $\br, \br^*$ viewed as independent complex variables, $\D R = |c_0 + c_1 \br^2 + c_2 (\br^2)^2 + ... |^2 \d^3\br \d^3\br^*$.

In general case $h(z) \neq const \cdot z$ integral also can be defined. Again, consideration goes through intermediate appearance of \dfuns . For that we make use of special structure of the exponential in (\ref{N-gamma-delta}) and temporarily pass to components of $\bv$ which remind spherical ones, but are modified for complex case,
\begin{equation}                                                                   %5
\bv = \rv \bn, ~~~ \rv = u + iw, ~~~ \bn = \beon \ch \rho + i \betw \sh \rho, ~~~ \beon^2 = 1 = \betw^2, ~~~ \beon \betw = 0.
\end{equation}

\noindent The orthogonal pair $\beon, \betw$ is specified by three angles, e. g. by azimuthal $\theta_1$ and polar $\varphi_1$ angles of $\beon$ and polar angle $\varphi_2$ of $\betw$ (in the plane orthogonal to $\beon$). The integration measure in the coordinates $u, w, \rho, \theta_1, \varphi_1, \varphi_2$
\begin{equation}                                                                   %6
\d^6 \bv \equiv \d^3 \Re \bv \d^3 \Im \bv = (u^2 + w^2)^2\d u \d w \d^4 \bn, ~~~ \d^4 \bn = \ch \rho \sh \rho \d \rho \sin \theta_1 \d \theta_1 \d \varphi_1 \d \varphi_2.
\end{equation}

\noindent Unlike the Euclidean case, $\bn$ varies in the noncompact region. Whenever this might violate convergence of some intermediate integrals over $\d^4 \bn$ below, we could imply some intermediate regularization being applied to these, e. g. $|\rho | < \Lambda$ at some large but finite $\Lambda$. The $u, w$ are defined via $(u + iw)^2 = \bv^2$, i. e. region of variation for $u + iw$ is a half of the complex plane. For example, for the standard choice of the cut for square root function $u \geq 0$. However, integration over $\d u$ in (\ref{Malbegade}) can be extended to the full real axis $(-\infty, +\infty)$. This is only possible because formal putting $u \to -u, w \to -w$ is equivalent to $\beon \to -\beon, \betw \to -\betw$ in (\ref{Malbegade}) due to the oddness of $h(z)$. Such identity of integration points leads to \dfuns of $h$,
\begin{eqnarray}\label{intvh}                                                      %7
 & & \int \exp \left [{i\over 2}\rv h(\bn\br) + {i\over 2}\rv^*h(\bn\br)^*\right ] (\bv^2)^l (\bv^{*2})^m \vaonajbonbk \d^6 \bv \nonumber\\ & & \hspace{-1cm} = {1\over 2} \int \naonajbonbk \d^4 \bn \int\limits^{+\infty}_{-\infty} \d u \int\limits^{+\infty}_{-\infty} \d w e^{i[uf(\bsn\bsr) + wg(\bsn\bsr)]}(u + iw)^{j+2l+2}(u - iw)^{k+2m+2} \nonumber\\ & & \hspace{-1cm} = {1\over 2} \int \naonajbonbk \d^4 \bn (2\pi)^2 \left ({\partial \over i\partial f} + {\partial \over \partial g}\right )^{j+2l+2} \left ({\partial \over i\partial f} - {\partial \over \partial g}\right )^{k+2m+2} \de (f) \de (g) \nonumber\\ & & = {1\over 2} \int \naonajbonbk \d^4 \bn (2\pi)^2 \left (\twi \ddh \right )^{j+2l+2} \left (\twi \ddhstar \right )^{k+2m+2} \detw (h)
\end{eqnarray}

\noindent where $f(z) = \Re h(z)$, $g(z) = -\Im h(z)$, $\de (f) \de (g) \equiv \detw (h)$,
\begin{equation}                                                                   %8
{\partial \over i\partial f} + {\partial \over \partial g} \equiv \twi \ddh, ~~~
{\partial \over i\partial f} - {\partial \over \partial g} \equiv \twi \ddhstar.
\end{equation}

\noindent Derivatives of $\detw (h)$ expand into combinations of the derivatives of $\detw (z) \equiv \de (x) \de (y)$, $z \equiv  \bn \br = x - iy$. These combinations can be found by applying to probe functions $\varphi (z, z^*)$,
\begin{eqnarray}\label{probe}                                                      %9
 & & \int \left [\left (\ddh\right )^{j+2l+2}\left (\ddhstar\right )^{k+2m+2}\detw (h)\right ] \varphi (z, z^*) \d^2 z \nonumber\\ & & \hspace{-1cm} = (-1)^{j+k}\left (\ddh \right )^{j+2l+2}\left (\ddhstar \right )^{k+2m+2} \left [\varphi (z(h), z(h)^*){\d z \over \d h}{\d z^* \over \d h^*}\right ]_{h,h^*=0}, ~~ \d^2 z = \d x \d y. \hspace{5mm}
\end{eqnarray}

\noindent Let us choose
\begin{equation}                                                                  %10
\varphi (z, z^*) = {\varphinp (0,0) \over n!p!} z^n z^{*p},
\end{equation}

\noindent thus we find coefficient of $\de^{(n)}(z)\de^{(p)}(z^*) \equiv (\d / \d z)^n (\d / \d z^*)^p \de (z) \de (z^*)$ (here $2\de (z) \de (z^*) \equiv \detw (z)$) in $\de^{(j+2l+2)}(h) \de^{(k+2m+2)}(h^*)$,
\begin{eqnarray}\label{delta-h-to-z}                                              %11
\de^{(j+2l+2)}(h) \de^{(k+2m+2)}(h^*) = ... + (-1)^{j+n+k+p} \de^{(n)}(z)\de^{(p)}(z^*) \nonumber\\ \cdot \left (\ddh \right )^{j+2l+3}\left (\ddhstar \right )^{k+2m+3} \left [ {z(h)^{n+1}\over (n+1)!}{z(h)^{*p+1}\over (p+1)!} \right ]_{h,h^*=0} + ... .
\end{eqnarray}

\noindent (This term is nonzero at $(j-n)({\rm mod}2) = 0$, $(k-p)({\rm mod}2) = 0$, $j + 2l + 2 \geq n$, $k + 2m + 2 \geq p$.) Let us apply formula (\ref{intvh}) read from right to left to appearing here $\de^{(n)}(z)\de^{(p)}(z^*)$, now for $h(z) = z$,
\begin{eqnarray}\label{delta-back}                                                %12
& & {1\over 2} \int \naonajbonbk \d^4 \bn (2\pi)^2 \left (\twi \ddz \right )^{n} \left (\twi \ddzstar \right )^{p} \detw (z) \nonumber\\ & & \hspace{-1cm} = {1\over 2} \int \naonajbonbk \d^4 \bn (2\pi)^2 \left ({\partial \over i\partial x} + {\partial \over \partial y}\right )^{n} \left ({\partial \over i\partial x} - {\partial \over \partial y}\right )^{p} \de (x) \de (y) \nonumber\\ & & \hspace{-1cm} = {1\over 2} \int \naonajbonbk \d^4 \bn \int\limits^{+\infty}_{-\infty} \d u \int\limits^{+\infty}_{-\infty} \d w e^{i[u \Re (\bsn\bsr) - w \Im (\bsn\bsr)]}(u + iw)^{n}(u - iw)^{p} \nonumber\\ & & = \int \exp \left ({i\over 2}\bv\br + {i\over 2}\bv^*\br^*\right ) \rv^{n-j} \rv^{*p-k} \vaonajbonbk {\d^6 \bv \over \rv^2 \rv^{*2}}.
\end{eqnarray}

\noindent Substitute (\ref{delta-h-to-z}) and (\ref{delta-back}) to (\ref{intvh}) and integrate (\ref{intvh}) over $\rconclamdondmu \D R$. We get for the moment
\begin{eqnarray}                                                                  %13
& & \Malbegade (l,m) = \int \rconclamdondmu \D R \int \exp \left [{i\over 2}\rv h(\bn\br) + {i\over 2}\rv^*h(\bn\br)^*\right ] \nonumber\\ & & \cdot (\bv^2)^l (\bv^{*2})^m \vaonajbonbk \d^6 \bv = \int \rconclamdondmu \D R \nonumber\\ & & \cdot (-1)^{j+k} \left ( \twi \ddh \right )^{j+2l+2} \left ( \twi \ddhstar \right )^{k+2m+2} {\d z \over \d h} {\d z^* \over \d h^*} \int \exp \left ({i\over 2}\bv\br + {i\over 2}\bv^*\br^*\right ) \nonumber\\ & & \cdot \vaonajvjbonbkvk {\d^6 \bv \over \rv^2 \rv^{*2}} \left \{ ... + {[\rv z/(2i)]^n \over n!} {[\rv^*z^*/(2i)]^p \over p!} + ... \right \}_{h,h^* = 0}.
\end{eqnarray}

\noindent We can extend summation over $n, p$ to that over infinite set of nonnegative integers of which only finite number of terms at the given finite $j, k, l, m$ (pointed out after formula (\ref{delta-h-to-z})) are active. Thus we have
\begin{equation}\label{I-to-M}                                                    %14
\Malbegade (l,m) = \left ( 2i \ddh \right )^{j+2l+2} \left ( 2i \ddhstar \right )^{k+2m+2} \left [ {\d z \over \d h} {\d z^* \over \d h^*} \Ialbegade (z,z^*) \right ]_{h,h^* = 0}
\end{equation}

\noindent where
\begin{eqnarray}\label{Ialbegade}                                                 %15
& & \Ialbegade (z,z^*) = \int \rconclamdondmu \D R \int \exp \left ({i\over 2}\bv\br + {i\over 2}\bv^*\br^*\right ) \vaonajvjbonbkvk \nonumber\\ & & \cdot \left [ {1\over 2} \exp {\rv z \over 2i} + {(-1)^j \over 2} \exp{i\rv z \over 2}\right ] \left [ {1\over 2} \exp {\rv^*z^* \over 2i} + {(-1)^k \over 2} \exp{i\rv^*z^* \over 2}\right ] {\d^6 \bv \over \rv^2 \rv^{*2}}
\end{eqnarray}

\noindent is "generating function".

\medskip

\noindent {\bf\large 3. Factorization into (anti-)selfdual parts}

\noindent At $\hzarcz$ such factorization for $\Ngade$ defined from moments (\ref{I-to-M}) can be proven. (Although the case $\hz$ could be treated in more simple way via intermediate appearance of vector \dfuns $\deth (\Re \br) \deth (\Im \br)$ as mentioned in the beginning of paragraph 2.)

Consider the terms like $z^nC_n(z^*)$ or $C_n(z)z^{*n}$ in $\Ialbegade (z,z^*)$ with holomorphic $C_n(z)$.

\medskip

\noindent LEMMA 1. {\it Adding $z^nC_n(z^*)$ or $C_n(z)z^{*n}$ at a nonnegative integer $n$ with $C_n(z)$ holomorphic at $z = 0$ to $\Ialbegade (z,z^*)$ does not contribute to $\Ngade (\bv, \bv^*)$ at $\harcz$ in the region with points $\rv^2 = 4\tilde{n}^2 (1 + i/ \gamma )^{-2}$, $\tilde{n} = n + 1, n - 1, ..., n({\rm mod}2) + 1$ excluded.}

\medskip

{\it Proof.} Consider adding the term $z^nC_n(z^*)$. The dependence of contribution to $\Malbegade (l,m)$ on $l$ decouples as
\begin{equation}\label{z-nC-n}                                                    %16
\left (2i \ddh \right )^{j+2l+2} \left ( {\d z \over \d h } z^n \right )_{h=0} = \left. \left (2i \ddh \right )^{j+2l+3} {z^{n+1} \over 2i(n+1)} \right |_{h=0}.
\end{equation}

\noindent At $z = \sin {h \over 1 + i/ \gamma }$ the power $z^{n+1}$ contains harmonics $\sin {\tilde{n} h \over 1 + i/ \gamma }$ or $\cos {\tilde{n} h \over 1 + i/ \gamma }$, $\tilde{n} = n + 1, n - 1, ..., n({\rm mod}2) + 1$, for even or odd $n$, respectively. Contribution to the moment from harmonic $\sin {\tilde{n} h \over 1 + i/ \gamma }$ or $\cos {\tilde{n} h \over 1 + i/ \gamma }$ is proportional to ${1\over 2i(n+1)} \left ( {2i\tilde{n} \over 1+i/\gamma} \ddh \right )^{j+2l+3} (\sin h)$ or ${1\over 2i(n+1)} \left ( {2i\tilde{n} \over 1+i/\gamma} \ddh \right )^{j+2l+3} (\cos h)$, respectively. Nonzero contribution to the moment follows for even or odd $j$, respectively, and the dependence on $l$ is proportional to $[2\tilde{n} / (1 + i/ \gamma )]^{2l}$. This corresponds to the singular term in the functional
\begin{equation}                                                                  %17
\Malbegade (f(\rv^2) g(\rv^2)^*) = \int \Ngade (\bv, \bv^*) f(\rv^2) g(\rv^2)^* \vaonajbonbk \d^6 \bv
\end{equation}

\noindent proportional to $f(4\tilde{n}^2 (1 + i/ \gamma )^{-2})$, that is, to probe function $f(\rv^2)$ taken at the point $4\tilde{n}^2 (1 + i/ \gamma )^{-2}$. We may define set of probe (anti-)holomorphic functions $f(\rv^2), g(\rv^2)^*$ to vanish at $\rv^2 = 4\tilde{n}^2 (1 + i/ \gamma )^{-2}$, $\tilde{n} = n + 1, n - 1, ..., n({\rm mod}2) + 1$ for a given finite $n$. These functions probe $\Ngade (\bv, \bv^*)$ in the region with points $\rv^2 = 4\tilde{n}^2 (1 + i/ \gamma )^{-2}$, $\tilde{n} = n + 1, n - 1, ..., n({\rm mod}2) + 1$ excluded. The functional $\Malbegade$ defined on these functions does not change upon adding the terms $z^nC_n(z^*)$ and $C_n(z)z^{*n}$ to $\Ialbegade$, and $\Ngade (\bv, \bv^*)$ recovered from $\Malbegade$ defined on these functions, i. e. in the region with points $\rv^2 = 4\tilde{n}^2 (1 + i/ \gamma )^{-2}$, $\tilde{n} = n + 1, n - 1, ..., n({\rm mod}2) + 1$ excluded, does not change too. \vspace{-11mm} \begin{flushright} $\spadesuit$ \end{flushright} \vspace{-1mm}

Now we would like to prove factorization of $\Ngade (\bv, \bv^*)$ at $\hzarcz$ into holomorphic and antiholomorphic parts in the region with the above nonphysical points (at $\harcz$) excluded. To find $\Ngade (\bv, \bv^*)$ with certain multiindices $\gamma, \delta$ it is sufficient to know $\Malbegade (l,m)$ at certain $\alpha, \beta$ with certain lengths of these multiindices $j, k$ (normally the same as lengths $\lambda, \mu$ of $\gamma, \delta$, respectively).

\medskip

\noindent LEMMA 2. {\it The $\Ngade (\bv, \bv^*)$ at $\hzarcz$ if recovered from moments factorizes into holomorphic and antiholomorphic parts in the region of its definition and with nonphysical points (at $\harcz$) $\rv^2 = 4\tilde{\jmath}^2 (1 + i/ \gamma )^{-2}$, $\tilde{\jmath} = j + 1, j - 1, ..., j({\rm mod}2) + 1$ and $\rv^2 = 4\tilde{k}^2 (1 + i/ \gamma )^{-2}$, $\tilde{k} = k + 1, k - 1, ..., k({\rm mod}2) + 1$ excluded.}

\medskip

{\it Proof.} Let us subtract from $\exp (\pm i\rv z/2)$ and from $\exp (\pm i\rv^*z^*/2)$ in square brackets in the formula (\ref{Ialbegade}) for $\Ialbegade (z,z^*)$ the first up to $\propto (\rv z)^j$ inclusive and the first up to $\propto (\rv^*z^*)^k$ inclusive terms of the Taylor expansions of these functions over $z$ and over $z^*$, respectively. At $\harcz$ use LEMMA 1. At $\hz$ note that, e. g., the term $z^nC_n(z^*)$ in $\Ialbegade$ at $n \leq j$ does not contribute to the moments ((\ref{z-nC-n}) vanishes). The $\Ialbegade (z,z^*)$ is replaced by
\begin{eqnarray}\label{tilIalbegade}                                              %18
& & \tilIalbegade (z,z^*) = \int \rconclamdondmu \D R \int \exp \left ({i\over 2}\bv\br + {i\over 2}\bv^*\br^*\right ) \vaonajvjbonbkvk \nonumber\\ & & \cdot \left [ {1\over 2} \exp {\rv z \over 2i} + {(-1)^j \over 2} \exp{i\rv z \over 2} - \sum^{[j/2]}_{n=0} {1 \over (j - 2n)! } \left ({\rv z \over 2i}\right )^{j-2n}\right ] \nonumber \\ & & \cdot \left [ {1\over 2} \exp {\rv^*z^* \over 2i} + {(-1)^k \over 2} \exp{i\rv^*z^* \over 2} - \sum^{[k/2]}_{p=0} {1 \over (k - 2p)! } \left ({\rv^*z^* \over 2i}\right )^{k-2p}\right ] {\d^6 \bv \over \rv^2 \rv^{*2}}.
\end{eqnarray}

\noindent This generating function defines moments of some $\tilNgade (\bv, \bv^*)$ which coincides with $\Ngade (\bv, \bv^*)$ in the region with nonphysical points (at $\harcz$) $\rv^2 = 4\tilde{\jmath}^2 (1 + i/ \gamma )^{-2}$, $\tilde{\jmath} = j + 1, j - 1, ..., j({\rm mod}2) + 1$ and $\rv^{*2} = 4\tilde{k}^2 (1 - i/ \gamma )^{-2}$ or $\rv^2 = 4\tilde{k}^2 (1 + i/ \gamma )^{-2}$, $\tilde{k} = k + 1, k - 1, ..., k({\rm mod}2) + 1$ excluded. Expansion over $z, z^*$ gives nonnegative powers of $\rv^2, \rv^{*2}$,
\begin{eqnarray}                                                                  %19
& & \tilIalbegade (z, z^*) = \left ( {z \over 2i} \right )^{j+2} \left ( {z^* \over 2i} \right )^{k+2} \int \rconclamdondmu \D R \int \exp \left ( {i\over 2} \bv \br + {i\over 2} \bv^* \br^*\right ) \nonumber \\ & & \cdot \left [ \sum^{\infty}_{n=0} {(\rv z/2i)^{2n} \over (j+2+2n)!}\right ] \left [ \sum^{\infty}_{p=0} {(\rv^*z^*/2i)^{2p} \over (k+2+2p)!}\right ] \vaonajbonbk \d^6 \bv.
\end{eqnarray}

\noindent Upon separating real and imaginary parts $\bv = \bu + i\bw$, $\br = \bs - i\bq$ each term transforms through intermediate appearance of \dfuns,
\begin{eqnarray}\label{factoriz}                                                  %20
& & \int \rconclamdondmu \D R \int \exp \left ( {i\over 2} \bv \br + {i\over 2} \bv^* \br^*\right ) (\bv^2)^n (\bv^{*2})^p \vaonajbonbk \d^6 \bv \nonumber \\ & & = \int \rconclamdondmu \D R \int \exp (i\bu\bs + i\bw\bq ) [(\bu + i \bw )^2]^n [(\bu - i \bw )^2]^p \nonumber \\ & & \cdot (u^{a_1} + iw^{a_1}) ... (u^{a_j} + iw^{a_j}) (u^{b_1} - iw^{b_1}) ... (u^{b_k} - iw^{b_k}) \d^3 \bu \d^3 \bw \nonumber \\ & & = \int \rconclamdondmu \D R \,\,\, (2\pi)^6 \left [ \left ( {\partial \over i\partial \bs} + {\partial \over \partial \bq} \right )^2 \right ]^n \left [ \left ( {\partial \over i\partial \bs} - {\partial \over \partial \bq} \right )^2 \right ]^p \nonumber \\ & & \hspace{-8mm} \cdot \left ( { \partial \over i\partial s_{a_1} } + { \partial \over \partial q_{a_1} } \right ) \dots \left ( { \partial \over i\partial s_{a_j} } + { \partial \over \partial q_{a_j} } \right ) \left ( { \partial \over i\partial s_{b_1} } - { \partial \over \partial q_{b_1} } \right ) \dots \left ( { \partial \over i\partial s_{b_k} } - { \partial \over \partial q_{b_k} } \right ) \deth (\bs ) \deth (\bq ) \nonumber \\ & & = 8\pi^2 \left \{ (2i)^{j+2n} {\partial \over \partial r_{a_1}} \dots {\partial \over \partial r_{a_j}} \left [ \left ( {\partial \over \partial \br} \right )^2 \right ]^n { r_{c_1} \dots r_{c_{\lambda}} \over \br^2} \left ( {1 \over \sqrt{1 - \br^2} } - 1 \right ) \right \}_{\br = 0} \nonumber \\ & & \cdot \left \{ (2i)^{k+2p} {\partial \over \partial r^*_{b_1}} \dots {\partial \over \partial r^*_{b_k}} \left [ \left ( {\partial \over \partial \br^* } \right )^2 \right ]^p { r^*_{d_1} \dots r^*_{d_{\mu}} \over \br^{*2} } \left ( {1 \over \sqrt{1 - \br^{*2} } } - 1 \right ) \right \}_{\br^* = 0}.
\end{eqnarray}

\noindent Here
\begin{equation}                                                                  %21
{\partial \over i\partial s_a} + {\partial \over \partial q_a} \equiv \twi { \partial \over \partial r_a }, ~~~
{\partial \over i\partial s_a} - {\partial \over \partial q_a} \equiv \twi { \partial \over \partial r^*_a }
\end{equation}

\noindent and
\begin{equation}                                                                  %22
{\partial \over \partial r^*_a } { r_{c_1} \dots r_{c_{\lambda}} \over \br^2} \left ( {1 \over \sqrt{1 - \br^2} } - 1 \right ) = 0, ~~~ {\partial \over \partial r_a } { r^*_{d_1} \dots r^*_{d_{\mu}} \over \br^{*2} } \left ( {1 \over \sqrt{1 - \br^{*2} } } - 1 \right ) = 0
\end{equation}

\noindent due to analyticity (Cauchy-Riemann conditions). This is key point for the factorization to occur. Complex dummy variables $\br, \br^*$ in the RHS of (\ref{factoriz}) can equally be viewed as real independent variables, the result being the same. This looks as possibility to replace integration over SO(3,1) by integration over SO(4). Eventually we trace back to (\ref{tilIalbegade}) where now $\bv, \br, z$ on one hand and $\bv^*, \br^*, z^*$ on another hand can be taken as independent real variables. Then $z, z^*$ on which the result $\tilIalbegade$ depends can be continued to the desired region. Therefore $2^3 \tilIalbegade (z, z^*) = \tilIalga (z) \tilIdotbedotde (z^*)$ where
\begin{eqnarray}%\label{tilIalga}                                                 %23
& & \tilIalga (z) = \int r_{c_1} \dots r_{c_{\lambda}} \left ( {1 \over \sqrt{1 - \br^2}} -1 \right ) {\d^3 \br \over 8\pi^2 \br^2} \int \exp \left ({i\over 2}\bv\br \right ) \nonumber \\ & & \cdot {v^{a_1} \dots v^{a_j} \over \rv^j} \left [ {1\over 2} \exp {\rv z \over 2i} + {(-1)^j \over 2} \exp{i\rv z \over 2} - \sum^{[j/2]}_{n=0} {1 \over (j - 2n)! } \left ({\rv z \over 2i}\right )^{j-2n}\right ] {\d^3 \bv \over \rv^2}.
\end{eqnarray}

\noindent Here integration is performed over real SO(3), $\Im \br =0, \br^2 \leq 1$, and over real $\bv$. Evidently, $\tilNgade (\bv, \bv^*)$ (that is $\Ngade (\bv, \bv^*)$ outside singularity points) restored from these $\tilIalbegade (z, z^*)$ should factorize too, $\tilNgade (\bv, \bv^* ) = \tilN_{\gamma} (\bv ) \tilN_{\dot{\delta}} (\bv^* )$. \vspace{-11mm} \begin{flushright} $\spadesuit$ \end{flushright} \vspace{-1mm}

\noindent A pleasant feature arising in this proof is correspondence with SO(4) (Euclidean) case.

\medskip

\noindent {\bf\large 4. The simplest (basic) integral}

\noindent Consider important particular case when $\Ngade = \N$ is scalar (indices $\gamma, \dot{\delta}$ are empty), the more general expressions have similar features. Then nontrivial $\tilIalga (z) \equiv \tilde{I}^{\alpha} (z), \tilIdotbedotde (z^*) \equiv \tilde{I}^{\dot{\beta}} (z^*)$ are expressible in the simplest way in terms of metric tensor $g^{ab}$ and scalars, i. e. it is sufficient to consider empty $\alpha, \dot{\beta}$ as well. This also means that $\tilN (\bv, \bv^*)$ can be considered as function of $\rv^2, \rv^{*2}$ only (although this is evident in this simple case from the very beginning since there is no singled out vector(s) on which structures over $\alpha, \dot{\beta}$ might depend). It will not cause confusion if we shall denote this function by the same symbol $\tilN (\rv^2, \rv^{*2})$. We have $\tilIalga (z) \equiv \tilde{I} (z), \tilIdotbedotde (z^*) \equiv \tilde{I} (z^*)$,
\begin{eqnarray}%\label{tilIalga}                                                 %24
& & \tilde{I} (z) = \int \left ( {1 \over \sqrt{1 - \br^2}} -1 \right ) {\d^3 \br \over 8\pi^2 \br^2} \int \exp \left ({i\over 2}\bv\br \right ) \left (\cos {\rv z \over 2} - 1 \right ) {\d^3 \bv \over \rv^2} \nonumber \\ & & = 2 \pi \ln { 1 + \sqrt{1 - z^2} \over 2}
\end{eqnarray}

\noindent and
\begin{eqnarray}\label{tilN}                                                      %25
& & \int \tilN (\rv^2, \rv^{*2}) \rv^{2l} \rv^{*2m} \d^6 \bv = 2^{-3} \tilde{N}(\rv^{2l}) \tilde{N}(\rv^{2m})^*, \nonumber \\ & & \tilde{N} (\rv^{2l}) = \pi (-1)^l \left (2 \ddh \right )^{2l+2} \left (2{\d z\over \d h} \ln {1 + \sqrt{1 - z^2} \over 2}\right )_{h=0} \nonumber \\ & & = \pi (-1)^l \left [{1\over 2}\left (1 + {i\over \gamma}\right )\right ]^{-2l-3} \left (\ddh \right )^{2l+2} \left (\cos h \ln {1 + \cos h \over 2}\right )_{h=0}
\end{eqnarray}

\noindent at $z = \sin {h \over 1 + i/ \gamma }$ (rescaling $h \to (1 + i / \gamma)h$ is made). Let us use the value of the following table integral,
\begin{equation}                                                                  %26
\int\limits^{\infty}_0 {l \over l^2 + 1} {\ch hl\over \sh \pi l} \d l = {h\over 2} \sin h - {1\over 2} + {1\over 2} \cos h \ln [2(1 + \cos h)],
\end{equation}

\noindent to express RHS of (\ref{tilN}) in terms of it and thus map differentiation over $h$ to operation of multiplication. The terms $\cos h$ and $h \sin h$ lead to appearance of the terms $\propto f(\rv^2)_{\rsv^2 = 4(1+i/\gamma)^{-2}}$ and $\propto f^{\prime}(\rv^2)_{\rsv^2 = 4(1+i/\gamma)^{-2}}$ in the functional $\tilde{N}(f(\rv^2))$. Situation is analogous to that one appeared in LEMMA 1, and the set of probe functions $f(\rv^2)$ as already chosen in LEMMA 2 for the considered scalar case $j = 0, k = 0$ is vanishing at the nonphysical point $\rv^2 = 4(1+i/\gamma)^{-2}$. Additional requirement is that also first derivatives of these functions be vanishing at this point. Upon simple transformation of integration contours the functional in question reads
\begin{eqnarray}\label{M}                                                         %27
2^3 \tilde{M}(f(\rv^2)g(\rv^2)^*) = \int \tilN (\rv^2, \rv^{*2}) f(\rv^2) g(\rv^2)^* 2^3 \d^6 \bv \nonumber \\ = {i\over 2} \int {(1/\gamma - i) \rv/2\over (1/\gamma - i)^2 \rv^2/4 + 1} {f(\rv^2) \d^3 \bv \over \sh [\pi (1/\gamma - i) \rv/2]} \nonumber \\ \cdot {-i\over 2} \int {(1/\gamma + i) \tilde{\rv}/2\over (1/\gamma + i)^2 \tilde{\rv}^2/4 + 1} {g^*(\tilde{\rv}^2) \d^3 \tilde{\bv} \over \sh [\pi (1/\gamma + i) \tilde{\rv}/2]}
\end{eqnarray}

\noindent at $f(4(1 + i/ \gamma )^{-2}) = 0$, $f^{\prime}(4(1 + i/ \gamma )^{-2}) = 0$, $g(4(1 + i/ \gamma )^{-2}) = 0$, $g^{\prime}(4(1 + i/ \gamma )^{-2}) = 0$; integrals in the RHS are over {\it independent real} $\bv, \tilde{\bv}$. We have introduced notation $g^* (\tilde{\rv}^2) = g(\tilde{\rv}^{2*})^* ( \equiv [g(\tilde{\rv}^{2*})]^* )$ having in view generalization to complex $\tilde{\rv}^2$. To pass to integration over complex $\bv, \bv^*$ we consider integrals in the RHS of (\ref{M}) as single 6-fold integral over $\d^3 \bv \d^3 \tilde{\bv}$, redenote $\bv = \bu + \bw$, $\tilde{\bv} = \bu - \bw$, and, for $\bw = \rw \rbn$, $\rbn^2 = 1$ rotate integration interval $\rw \in (0, \infty )$ in the plane of complex $\rw$ according to $\rw \to i\rw$. Thus we arrive at the desired form modulo pole contribution of the terms proportional to $f(\rv^2)$ or/and $g(\rv^2)^*$ taken at nonphysical points $\rv^2 = 4n^2(1 + i/\gamma )^{-2}$. So we get\footnote{At $\hz$ we reproduce (the module squared of) our result \citeup{Kha2} on 3 dimensional SO(3) gravity $Ki_1 (l) (2\pi l)^{-1}$ continued to complex argument $l = \sqrt{(1/\gamma - i)^2\rv^2}/2$ (the branch of square root is chosen in standard way such that $\Re \sqrt{(1/\gamma - i)^2\rv^2} \geq 0$), with modified integral Bessel function $Ki_1 (l)$ which exponentially decays at $\Re l \to \infty$.}
\begin{equation}\label{Nvv}                                                       %28
\N (\rv^2, \rv^{*2}) = \left | {1 \over {1 \over 4}\left ({1\over \gamma} - i\right )^2 \rv^2 + 1} \cdot {{1 \over 4}\left ({1\over \gamma} - i\right ) \rv \over \sh \left [{\pi \over 2} \left ({1\over \gamma} - i\right ) \rv \right ]} \right |^2
\end{equation}

\noindent in the region with the points $\rv^2 = 4n^2(1 + i/\gamma)^{-2}, n = 1, 2, ...$ excluded.

The form of dependence providing singularity of $\N (\rv^2, \rv^{*2})$ (\ref{Nvv}) at $\rv^2\!$ $\!= 4n^2(1 + i/\gamma)^{-2}, n\!$ $\!= 1, 2, ...$ looks as result of a summation in the path integral over branches of the 'arcsin' function, as if we had substituted $\arcsin \to \arcsin + 2\pi n$ in the exponential and summed over $n$ for each of the two 'arcsin' functions. (This would just result in the hyperbolic or trigonometric function in the denominator of (\ref{Nvv}).) Thus, when calculating any moment of distribution $\N$ we have dealt with only the values of a finite number of derivatives i. e. with {\it local} properties of the {\it principal} value of $\arcsin z$ at $z = 0$. Nevertheless when restoring $\N$ from the moments we have recovered full non-perturbative picture.

\begin{figure}[h]
\setlength{\unitlength}{1pt}
\begin{picture}(300,306)(-210,-10)

\multiput(0,0)(0,1){300}{.}
\multiput(70,0)(0,1){300}{.}
\multiput(0,0)(-1,0){200}{.}
\multiput(70,0)(1,0){150}{.}
\multiput(70,0)(-1,0){20}{.}

\put(2,-10){0}
\put(15,-10){\scriptsize{$\rv^2\gamma^{-2}/4$}}
\put(66,-10){0}
\put(205,-10){\scriptsize{$\rv^2/4$}}

\multiput(0,150)(0.5,0){10}{.}
\multiput(70,150)(-0.5,0){10}{.}

\put(7,147){10}
\put(53,147){10}

\put(3,290){\scriptsize{$(2\pi)^2\N$}}
\put(41,290){\scriptsize{$(2\pi)^2\N$}}

\put(-100,-25){$\gamma=0.05$}
\put(115,-25){$\gamma=10$}

%frameHeight=300.000000 %A20=-0.100000 %frameStep=1.000000 %gamma=0.050000 %A2max=0.020000 %horizontalFrameToA2Scale=2000.000000 %verticalFrameTo4pi2Nscale=15.000000 %horizontalFrameShift=0.000000

\put(-200.0,1.9){.} \put(-199.0,1.9){.} \put(-198.0,2.0){.} \put(-197.0,2.1){.} \put(-196.0,2.1){.} \put(-195.0,2.2){.} \put(-194.0,2.3){.} \put(-193.0,2.4){.} \put(-192.0,2.5){.} \put(-191.0,2.6){.} \put(-190.0,2.7){.} \put(-189.0,2.8){.} \put(-188.0,2.9){.} \put(-187.1,3.0){.} \put(-186.1,3.1){.} \put(-185.1,3.3){.} \put(-184.1,3.4){.} \put(-183.1,3.4){.} \put(-182.1,3.5){.} \put(-181.1,3.6){.} \put(-180.1,3.7){.} \put(-179.1,3.7){.} \put(-178.1,3.7){.} \put(-177.1,3.7){.}

\multiput(-177.1,-3)(0,0.5){6}{.} %max %fourpi2N=0.248649 %A2=-0.088543 %A1=0.297562

\put(-185.1,-10){\scriptsize{$-36$}}

\put(-176.1,3.7){.} \put(-175.1,3.7){.} \put(-174.1,3.7){.} \put(-173.1,3.6){.} \put(-172.1,3.5){.} \put(-171.1,3.5){.} \put(-170.1,3.4){.} \put(-169.1,3.3){.} \put(-168.1,3.2){.} \put(-167.1,3.2){.} \put(-166.1,3.1){.} \put(-165.1,3.0){.} \put(-164.1,3.0){.} \put(-163.1,2.9){.} \put(-162.1,2.8){.} \put(-161.1,2.8){.} \put(-160.1,2.7){.} \put(-159.1,2.7){.} \put(-158.1,2.7){.} \put(-157.1,2.6){.} \put(-156.1,2.6){.} \put(-155.1,2.6){.}

%\multiput(-155.1,-3)(0,0.5){6}{.} %min %fourpi2N=0.175069 %A2=-0.077560 %A1=0.278495

\put(-154.1,2.6){.} \put(-153.1,2.6){.} \put(-152.1,2.7){.} \put(-151.1,2.7){.} \put(-150.1,2.7){.} \put(-149.1,2.8){.} \put(-148.1,2.8){.} \put(-147.1,2.9){.} \put(-146.1,3.0){.} \put(-145.1,3.1){.} \put(-144.1,3.2){.} \put(-143.1,3.3){.} \put(-142.2,3.4){.} \put(-141.2,3.6){.} \put(-140.2,3.8){.} \put(-139.2,4.0){.} \put(-138.2,4.2){.} \put(-137.2,4.4){.} \put(-136.3,4.7){.} \put(-135.3,5.0){.} \put(-134.4,5.3){.} \put(-133.4,5.6){.} \put(-132.5,6.0){.} \put(-131.5,6.3){.} \put(-130.6,6.7){.} \put(-129.7,7.1){.} \put(-128.7,7.4){.} \put(-127.8,7.8){.} \put(-126.9,8.1){.} \put(-125.9,8.3){.} \put(-124.9,8.5){.} \put(-123.9,8.6){.} \put(-122.9,8.6){.}

\multiput(-122.9,-3)(0,0.5){6}{.} %max %fourpi2N=0.574708 %A2=-0.061472 %A1=0.247934

\put(-130.9,-10){\scriptsize{$-25$}}

\put(-121.9,8.6){.} \put(-120.9,8.4){.} \put(-120.0,8.2){.} \put(-119.0,7.9){.} \put(-118.0,7.7){.} \put(-117.1,7.4){.} \put(-116.1,7.1){.} \put(-115.2,6.8){.} \put(-114.2,6.5){.} \put(-113.2,6.2){.} \put(-112.3,6.0){.} \put(-111.3,5.8){.} \put(-110.3,5.6){.} \put(-109.3,5.4){.} \put(-108.4,5.3){.} \put(-107.4,5.2){.} \put(-106.4,5.1){.} \put(-105.4,5.0){.} \put(-104.4,5.0){.} \put(-103.4,5.0){.}

%\multiput(-103.4,-3)(0,0.5){6}{.} %min %fourpi2N=0.332248 %A2=-0.051684 %A1=0.227341

\put(-102.4,5.0){.} \put(-101.4,5.1){.} \put(-100.4,5.2){.} \put(-99.4,5.3){.} \put(-98.4,5.4){.} \put(-97.4,5.6){.} \put(-96.4,5.9){.} \put(-95.5,6.2){.} \put(-94.5,6.5){.} \put(-93.6,6.9){.} \put(-92.7,7.4){.} \put(-91.8,7.9){.} \put(-91.0,8.5){.} \put(-90.2,9.1){.} \put(-89.4,9.8){.} \put(-88.7,10.6){.} \put(-88.0,11.3){.} \put(-87.4,12.1){.} \put(-86.7,13.0){.} \put(-86.2,13.8){.} \put(-85.6,14.7){.} \put(-85.1,15.5){.} \put(-84.6,16.4){.} \put(-84.1,17.3){.} \put(-83.6,18.2){.} \put(-83.1,19.1){.} \put(-82.6,19.9){.} \put(-82.1,20.8){.} \put(-81.6,21.6){.} \put(-81.0,22.3){.} \put(-80.4,23.0){.} \put(-79.7,23.5){.} \put(-78.8,23.7){.}

\multiput(-78.8,-3)(0,0.5){6}{.} %max %fourpi2N=1.577670 %A2=-0.039385 %A1=0.198457

\put(-86.8,-10){\scriptsize{$-16$}}

\put(-77.8,23.2){.} \put(-77.0,22.5){.} \put(-76.3,21.7){.} \put(-75.7,20.8){.} \put(-75.1,20.0){.} \put(-74.6,19.2){.} \put(-74.0,18.3){.} \put(-73.4,17.5){.} \put(-72.9,16.7){.} \put(-72.3,15.9){.} \put(-71.7,15.2){.} \put(-71.0,14.4){.} \put(-70.4,13.7){.} \put(-69.7,13.1){.} \put(-68.9,12.5){.} \put(-68.1,11.9){.} \put(-67.2,11.4){.} \put(-66.3,11.0){.} \put(-65.4,10.7){.} \put(-64.4,10.5){.} \put(-63.5,10.5){.}

%\multiput(-63.5,-3)(0,0.5){6}{.} %min %fourpi2N=0.697287 %A2=-0.031729 %A1=0.178128

\put(-62.5,10.5){.} \put(-61.5,10.6){.} \put(-60.5,10.9){.} \put(-59.5,11.3){.} \put(-58.7,11.9){.} \put(-57.8,12.5){.} \put(-57.1,13.2){.} \put(-56.4,14.0){.} \put(-55.8,14.9){.} \put(-55.2,15.8){.} \put(-54.7,16.7){.} \put(-54.3,17.7){.} \put(-53.9,18.6){.} \put(-53.5,19.6){.} \put(-53.2,20.6){.} \put(-52.8,21.6){.} \put(-52.6,22.6){.} \put(-52.3,23.6){.} \put(-52.0,24.6){.} \put(-51.8,25.6){.} \put(-51.6,26.6){.} \put(-51.4,27.6){.} \put(-51.2,28.6){.} \put(-51.0,29.6){.} \put(-50.8,30.6){.} \put(-50.6,31.6){.} \put(-50.5,32.7){.} \put(-50.3,33.7){.} \put(-50.2,34.7){.} \put(-50.0,35.7){.} \put(-49.9,36.7){.} \put(-49.8,37.7){.} \put(-49.6,38.7){.} \put(-49.5,39.7){.} \put(-49.4,40.7){.} \put(-49.3,41.7){.} \put(-49.2,42.7){.} \put(-49.1,43.8){.} \put(-49.0,44.8){.} \put(-48.9,45.8){.} \put(-48.7,46.8){.} \put(-48.6,47.8){.} \put(-48.6,48.8){.} \put(-48.5,49.8){.} \put(-48.4,50.8){.} \put(-48.3,51.8){.} \put(-48.2,52.8){.} \put(-48.1,53.8){.} \put(-48.0,54.8){.} \put(-47.9,55.8){.} \put(-47.8,56.8){.} \put(-47.7,57.8){.} \put(-47.7,58.8){.} \put(-47.6,59.8){.} \put(-47.5,60.8){.} \put(-47.4,61.8){.} \put(-47.3,62.8){.} \put(-47.3,63.8){.} \put(-47.2,64.8){.} \put(-47.1,65.8){.} \put(-47.0,66.8){.} \put(-46.9,67.8){.} \put(-46.9,68.8){.} \put(-46.8,69.8){.} \put(-46.7,70.8){.} \put(-46.6,71.8){.} \put(-46.5,72.8){.} \put(-46.5,73.8){.} \put(-46.4,74.8){.} \put(-46.3,75.7){.} \put(-46.2,76.7){.} \put(-46.1,77.7){.} \put(-46.0,78.7){.} \put(-45.9,79.7){.} \put(-45.8,80.6){.} \put(-45.7,81.6){.} \put(-45.6,82.6){.} \put(-45.5,83.5){.} \put(-45.4,84.5){.} \put(-45.2,85.4){.} \put(-45.1,86.3){.} \put(-44.9,87.1){.} \put(-44.6,87.8){.}

\multiput(-44.6,-3)(0,0.5){6}{.} %max %fourpi2N=5.855414 %A2=-0.022316 %A1=0.149386

\put(-52.6,-10){\scriptsize{$-9$}}

\put(-44.1,87.5){.} \put(-43.7,86.1){.} \put(-43.6,85.0){.} \put(-43.4,84.0){.} \put(-43.3,82.9){.} \put(-43.1,81.9){.} \put(-43.0,80.9){.} \put(-42.9,79.9){.} \put(-42.8,78.9){.} \put(-42.7,77.9){.} \put(-42.6,76.9){.} \put(-42.5,75.8){.} \put(-42.4,74.8){.} \put(-42.3,73.8){.} \put(-42.2,72.8){.} \put(-42.1,71.8){.} \put(-42.0,70.8){.} \put(-41.9,69.9){.} \put(-41.8,68.9){.} \put(-41.8,67.9){.} \put(-41.7,66.9){.} \put(-41.6,65.9){.} \put(-41.5,64.9){.} \put(-41.4,63.9){.} \put(-41.3,62.9){.} \put(-41.2,61.9){.} \put(-41.1,60.9){.} \put(-41.0,59.9){.} \put(-40.9,58.9){.} \put(-40.8,58.0){.} \put(-40.7,57.0){.} \put(-40.6,56.0){.} \put(-40.5,55.0){.} \put(-40.4,54.0){.} \put(-40.3,53.0){.} \put(-40.2,52.1){.} \put(-40.0,51.1){.} \put(-39.9,50.1){.} \put(-39.8,49.1){.} \put(-39.7,48.1){.} \put(-39.5,47.2){.} \put(-39.4,46.2){.} \put(-39.3,45.2){.} \put(-39.1,44.3){.} \put(-39.0,43.3){.} \put(-38.8,42.3){.} \put(-38.7,41.4){.} \put(-38.5,40.4){.} \put(-38.3,39.4){.} \put(-38.1,38.5){.} \put(-37.9,37.5){.} \put(-37.7,36.6){.} \put(-37.5,35.6){.} \put(-37.2,34.7){.} \put(-37.0,33.8){.} \put(-36.7,32.9){.} \put(-36.4,32.0){.} \put(-36.1,31.1){.} \put(-35.7,30.2){.} \put(-35.3,29.4){.} \put(-34.7,28.6){.} \put(-34.1,28.0){.} \put(-33.4,27.6){.}

%\multiput(-33.4,-3)(0,0.5){6}{.} %min %fourpi2N=1.838502 %A2=-0.016684 %A1=0.129166

\put(-32.4,27.7){.} \put(-31.5,28.4){.} \put(-30.8,29.4){.} \put(-30.4,30.4){.} \put(-30.0,31.4){.} \put(-29.6,32.4){.} \put(-29.3,33.4){.} \put(-29.1,34.4){.} \put(-28.8,35.4){.} \put(-28.6,36.5){.} \put(-28.4,37.5){.} \put(-28.2,38.5){.} \put(-28.0,39.5){.} \put(-27.9,40.5){.} \put(-27.7,41.6){.} \put(-27.6,42.6){.} \put(-27.5,43.6){.} \put(-27.3,44.6){.} \put(-27.2,45.6){.} \put(-27.1,46.7){.} \put(-27.0,47.7){.} \put(-26.9,48.7){.} \put(-26.8,49.7){.} \put(-26.7,50.7){.} \put(-26.6,51.7){.} \put(-26.5,52.8){.} \put(-26.4,53.8){.} \put(-26.3,54.8){.} \put(-26.3,55.8){.} \put(-26.2,56.8){.} \put(-26.1,57.8){.} \put(-26.0,58.8){.} \put(-26.0,59.9){.} \put(-25.9,60.9){.} \put(-25.8,61.9){.} \put(-25.8,62.9){.} \put(-25.7,63.9){.} \put(-25.7,64.9){.} \put(-25.6,65.9){.} \put(-25.5,66.9){.} \put(-25.5,68.0){.} \put(-25.4,69.0){.} \put(-25.4,70.0){.} \put(-25.3,71.0){.} \put(-25.3,72.0){.} \put(-25.2,73.0){.} \put(-25.2,74.0){.} \put(-25.2,75.0){.} \put(-25.1,76.0){.} \put(-25.1,77.1){.} \put(-25.0,78.1){.} \put(-25.0,79.1){.} \put(-24.9,80.1){.} \put(-24.9,81.1){.} \put(-24.9,82.1){.} \put(-24.8,83.1){.} \put(-24.8,84.1){.} \put(-24.7,85.1){.} \put(-24.7,86.1){.} \put(-24.7,87.1){.} \put(-24.6,88.2){.} \put(-24.6,89.2){.} \put(-24.6,90.2){.} \put(-24.5,91.2){.} \put(-24.5,92.2){.} \put(-24.5,93.2){.} \put(-24.4,94.2){.} \put(-24.4,95.2){.} \put(-24.4,96.2){.} \put(-24.4,97.2){.} \put(-24.3,98.2){.} \put(-24.3,99.2){.} \put(-24.3,100.3){.} \put(-24.2,101.3){.} \put(-24.2,102.3){.} \put(-24.2,103.3){.} \put(-24.2,104.3){.} \put(-24.1,105.3){.} \put(-24.1,106.3){.} \put(-24.1,107.3){.} \put(-24.1,108.3){.} \put(-24.0,109.3){.} \put(-24.0,110.3){.} \put(-24.0,111.3){.} \put(-24.0,112.3){.} \put(-23.9,113.3){.} \put(-23.9,114.4){.} \put(-23.9,115.4){.} \put(-23.9,116.4){.} \put(-23.8,117.4){.} \put(-23.8,118.4){.} \put(-23.8,119.4){.} \put(-23.8,120.4){.} \put(-23.8,121.4){.} \put(-23.7,122.4){.} \put(-23.7,123.4){.} \put(-23.7,124.4){.} \put(-23.7,125.4){.} \put(-23.7,126.4){.} \put(-23.6,127.4){.} \put(-23.6,128.4){.} \put(-23.6,129.4){.} \put(-23.6,130.4){.} \put(-23.6,131.5){.} \put(-23.5,132.5){.} \put(-23.5,133.5){.} \put(-23.5,134.5){.} \put(-23.5,135.5){.} \put(-23.5,136.5){.} \put(-23.5,137.5){.} \put(-23.4,138.5){.} \put(-23.4,139.5){.} \put(-23.4,140.5){.} \put(-23.4,141.5){.} \put(-23.4,142.5){.} \put(-23.4,143.5){.} \put(-23.3,144.5){.} \put(-23.3,145.5){.} \put(-23.3,146.5){.} \put(-23.3,147.5){.} \put(-23.3,148.5){.} \put(-23.3,149.5){.} \put(-23.2,150.5){.} \put(-23.2,151.6){.} \put(-23.2,152.6){.} \put(-23.2,153.6){.} \put(-23.2,154.6){.} \put(-23.2,155.6){.} \put(-23.2,156.6){.} \put(-23.1,157.6){.} \put(-23.1,158.6){.} \put(-23.1,159.6){.} \put(-23.1,160.6){.} \put(-23.1,161.6){.} \put(-23.1,162.6){.} \put(-23.1,163.6){.} \put(-23.0,164.6){.} \put(-23.0,165.6){.} \put(-23.0,166.6){.} \put(-23.0,167.6){.} \put(-23.0,168.6){.} \put(-23.0,169.6){.} \put(-23.0,170.6){.} \put(-23.0,171.6){.} \put(-22.9,172.6){.} \put(-22.9,173.6){.} \put(-22.9,174.6){.} \put(-22.9,175.7){.} \put(-22.9,176.7){.} \put(-22.9,177.7){.} \put(-22.9,178.7){.} \put(-22.9,179.7){.} \put(-22.8,180.7){.} \put(-22.8,181.7){.} \put(-22.8,182.7){.} \put(-22.8,183.7){.} \put(-22.8,184.7){.} \put(-22.8,185.7){.} \put(-22.8,186.7){.} \put(-22.8,187.7){.} \put(-22.7,188.7){.} \put(-22.7,189.7){.} \put(-22.7,190.7){.} \put(-22.7,191.7){.} \put(-22.7,192.7){.} \put(-22.7,193.7){.} \put(-22.7,194.7){.} \put(-22.7,195.7){.} \put(-22.7,196.7){.} \put(-22.7,197.7){.} \put(-22.6,198.7){.} \put(-22.6,199.7){.} \put(-22.6,200.7){.} \put(-22.6,201.7){.} \put(-22.6,202.7){.} \put(-22.6,203.7){.} \put(-22.6,204.8){.} \put(-22.6,205.8){.} \put(-22.6,206.8){.} \put(-22.5,207.8){.} \put(-22.5,208.8){.} \put(-22.5,209.8){.} \put(-22.5,210.8){.} \put(-22.5,211.8){.} \put(-22.5,212.8){.} \put(-22.5,213.8){.} \put(-22.5,214.8){.} \put(-22.5,215.8){.} \put(-22.5,216.8){.} \put(-22.5,217.8){.} \put(-22.4,218.8){.} \put(-22.4,219.8){.} \put(-22.4,220.8){.} \put(-22.4,221.8){.} \put(-22.4,222.8){.} \put(-22.4,223.8){.} \put(-22.4,224.8){.} \put(-22.4,225.8){.} \put(-22.4,226.8){.} \put(-22.4,227.8){.} \put(-22.4,228.8){.} \put(-22.3,229.8){.} \put(-22.3,230.8){.} \put(-22.3,231.8){.} \put(-22.3,232.8){.} \put(-22.3,233.8){.} \put(-22.3,234.8){.} \put(-22.3,235.8){.} \put(-22.3,236.8){.} \put(-22.3,237.8){.} \put(-22.3,238.8){.} \put(-22.3,239.9){.} \put(-22.3,240.9){.} \put(-22.2,241.9){.} \put(-22.2,242.9){.} \put(-22.2,243.9){.} \put(-22.2,244.9){.} \put(-22.2,245.9){.} \put(-22.2,246.9){.} \put(-22.2,247.9){.} \put(-22.2,248.9){.} \put(-22.2,249.9){.} \put(-22.2,250.9){.} \put(-22.2,251.9){.} \put(-22.2,252.9){.} \put(-22.2,253.9){.} \put(-22.1,254.9){.} \put(-22.1,255.9){.} \put(-22.1,256.9){.} \put(-22.1,257.9){.} \put(-22.1,258.9){.} \put(-22.1,259.9){.} \put(-22.1,260.9){.} \put(-22.1,261.9){.} \put(-22.1,262.9){.} \put(-22.1,263.9){.} \put(-22.1,264.9){.} \put(-22.1,265.9){.} \put(-22.1,266.9){.} \put(-22.0,267.9){.} \put(-22.0,268.9){.} \put(-22.0,269.9){.} \put(-22.0,270.9){.} \put(-22.0,271.9){.} \put(-22.0,272.9){.} \put(-22.0,273.9){.} \put(-22.0,274.9){.} \put(-22.0,275.9){.} \put(-22.0,276.9){.} \put(-22.0,277.9){.} \put(-22.0,278.9){.} \put(-22.0,279.9){.} \put(-22.0,280.9){.} \put(-21.9,281.9){.} \put(-21.9,282.9){.} \put(-21.9,283.9){.} \put(-21.9,285.0){.} \put(-21.9,286.0){.} \put(-21.9,287.0){.} \put(-21.9,288.0){.} \put(-21.9,289.0){.} \put(-21.9,290.0){.} \put(-21.9,291.0){.} \put(-21.9,292.0){.} \put(-21.9,293.0){.} \put(-21.9,294.0){.} \put(-21.9,295.0){.} \put(-21.9,296.0){.} \put(-21.8,297.0){.} \put(-21.8,298.0){.} \put(-21.8,299.0){.} \put(-21.8,300.0){.}

\multiput(-19.7,-3)(0,0.5){6}{.} %max %fourpi2N=43.448068 %A2=-0.009872 %A1=0.099359

\put(-27.7,-10){\scriptsize{$-4$}}

\put(-17.2,299.5){.} \put(-17.2,298.5){.} \put(-17.2,297.5){.} \put(-17.2,296.5){.} \put(-17.1,295.5){.} \put(-17.1,294.5){.} \put(-17.1,293.5){.} \put(-17.1,292.5){.} \put(-17.1,291.5){.} \put(-17.1,290.5){.} \put(-17.1,289.5){.} \put(-17.1,288.5){.} \put(-17.1,287.5){.} \put(-17.1,286.5){.} \put(-17.0,285.5){.} \put(-17.0,284.5){.} \put(-17.0,283.5){.} \put(-17.0,282.5){.} \put(-17.0,281.5){.} \put(-17.0,280.5){.} \put(-17.0,279.5){.} \put(-17.0,278.5){.} \put(-17.0,277.5){.} \put(-17.0,276.5){.} \put(-16.9,275.5){.} \put(-16.9,274.5){.} \put(-16.9,273.5){.} \put(-16.9,272.5){.} \put(-16.9,271.5){.} \put(-16.9,270.6){.} \put(-16.9,269.6){.} \put(-16.9,268.6){.} \put(-16.9,267.6){.} \put(-16.8,266.6){.} \put(-16.8,265.6){.} \put(-16.8,264.6){.} \put(-16.8,263.6){.} \put(-16.8,262.6){.} \put(-16.8,261.6){.} \put(-16.8,260.6){.} \put(-16.8,259.6){.} \put(-16.7,258.6){.} \put(-16.7,257.6){.} \put(-16.7,256.6){.} \put(-16.7,255.6){.} \put(-16.7,254.6){.} \put(-16.7,253.6){.} \put(-16.7,252.6){.} \put(-16.7,251.6){.} \put(-16.6,250.6){.} \put(-16.6,249.6){.} \put(-16.6,248.6){.} \put(-16.6,247.6){.} \put(-16.6,246.6){.} \put(-16.6,245.6){.} \put(-16.6,244.7){.} \put(-16.6,243.7){.} \put(-16.5,242.7){.} \put(-16.5,241.7){.} \put(-16.5,240.7){.} \put(-16.5,239.7){.} \put(-16.5,238.7){.} \put(-16.5,237.7){.} \put(-16.5,236.7){.} \put(-16.4,235.7){.} \put(-16.4,234.7){.} \put(-16.4,233.7){.} \put(-16.4,232.7){.} \put(-16.4,231.7){.} \put(-16.4,230.7){.} \put(-16.3,229.7){.} \put(-16.3,228.7){.} \put(-16.3,227.7){.} \put(-16.3,226.7){.} \put(-16.3,225.7){.} \put(-16.3,224.7){.} \put(-16.2,223.8){.} \put(-16.2,222.8){.} \put(-16.2,221.8){.} \put(-16.2,220.8){.} \put(-16.2,219.8){.} \put(-16.2,218.8){.} \put(-16.1,217.8){.} \put(-16.1,216.8){.} \put(-16.1,215.8){.} \put(-16.1,214.8){.} \put(-16.1,213.8){.} \put(-16.0,212.8){.} \put(-16.0,211.8){.} \put(-16.0,210.8){.} \put(-16.0,209.8){.} \put(-16.0,208.8){.} \put(-16.0,207.9){.} \put(-15.9,206.9){.} \put(-15.9,205.9){.} \put(-15.9,204.9){.} \put(-15.9,203.9){.} \put(-15.8,202.9){.} \put(-15.8,201.9){.} \put(-15.8,200.9){.} \put(-15.8,199.9){.} \put(-15.8,198.9){.} \put(-15.7,197.9){.} \put(-15.7,196.9){.} \put(-15.7,195.9){.} \put(-15.7,194.9){.} \put(-15.6,194.0){.} \put(-15.6,193.0){.} \put(-15.6,192.0){.} \put(-15.6,191.0){.} \put(-15.5,190.0){.} \put(-15.5,189.0){.} \put(-15.5,188.0){.} \put(-15.5,187.0){.} \put(-15.4,186.0){.} \put(-15.4,185.0){.} \put(-15.4,184.1){.} \put(-15.3,183.1){.} \put(-15.3,182.1){.} \put(-15.3,181.1){.} \put(-15.2,180.1){.} \put(-15.2,179.1){.} \put(-15.2,178.1){.} \put(-15.1,177.1){.} \put(-15.1,176.2){.} \put(-15.1,175.2){.} \put(-15.0,174.2){.} \put(-15.0,173.2){.} \put(-14.9,172.2){.} \put(-14.9,171.2){.} \put(-14.9,170.3){.} \put(-14.8,169.3){.} \put(-14.8,168.3){.} \put(-14.7,167.3){.} \put(-14.7,166.3){.} \put(-14.6,165.4){.} \put(-14.6,164.4){.} \put(-14.5,163.4){.} \put(-14.4,162.5){.} \put(-14.4,161.5){.} \put(-14.3,160.5){.} \put(-14.2,159.6){.} \put(-14.1,158.6){.} \put(-14.1,157.7){.} \put(-13.9,156.8){.} \put(-13.8,156.0){.} \put(-13.6,155.2){.}

%\multiput(-13.6,-3)(0,0.5){6}{.} %min %fourpi2N=10.349471 %A2=-0.006810 %A1=0.082521

\put(-13.3,155.3){.} \put(-12.9,157.1){.} \put(-12.8,158.2){.} \put(-12.7,159.3){.} \put(-12.6,160.3){.} \put(-12.5,161.4){.} \put(-12.5,162.4){.} \put(-12.4,163.4){.} \put(-12.3,164.5){.} \put(-12.3,165.5){.} \put(-12.2,166.5){.} \put(-12.2,167.5){.} \put(-12.1,168.6){.} \put(-12.1,169.6){.} \put(-12.0,170.6){.} \put(-12.0,171.6){.} \put(-12.0,172.6){.} \put(-11.9,173.7){.} \put(-11.9,174.7){.} \put(-11.9,175.7){.} \put(-11.8,176.7){.} \put(-11.8,177.7){.} \put(-11.7,178.7){.} \put(-11.7,179.7){.} \put(-11.7,180.8){.} \put(-11.7,181.8){.} \put(-11.6,182.8){.} \put(-11.6,183.8){.} \put(-11.6,184.8){.} \put(-11.5,185.8){.} \put(-11.5,186.8){.} \put(-11.5,187.8){.} \put(-11.5,188.9){.} \put(-11.4,189.9){.} \put(-11.4,190.9){.} \put(-11.4,191.9){.} \put(-11.4,192.9){.} \put(-11.3,193.9){.} \put(-11.3,194.9){.} \put(-11.3,195.9){.} \put(-11.3,196.9){.} \put(-11.2,197.9){.} \put(-11.2,198.9){.} \put(-11.2,199.9){.} \put(-11.2,201.0){.} \put(-11.2,202.0){.} \put(-11.1,203.0){.} \put(-11.1,204.0){.} \put(-11.1,205.0){.} \put(-11.1,206.0){.} \put(-11.1,207.0){.} \put(-11.0,208.0){.} \put(-11.0,209.0){.} \put(-11.0,210.0){.} \put(-11.0,211.0){.} \put(-11.0,212.0){.} \put(-11.0,213.0){.} \put(-10.9,214.0){.} \put(-10.9,215.1){.} \put(-10.9,216.1){.} \put(-10.9,217.1){.} \put(-10.9,218.1){.} \put(-10.9,219.1){.} \put(-10.8,220.1){.} \put(-10.8,221.1){.} \put(-10.8,222.1){.} \put(-10.8,223.1){.} \put(-10.8,224.1){.} \put(-10.8,225.1){.} \put(-10.7,226.1){.} \put(-10.7,227.1){.} \put(-10.7,228.1){.} \put(-10.7,229.1){.} \put(-10.7,230.1){.} \put(-10.7,231.1){.} \put(-10.7,232.2){.} \put(-10.6,233.2){.} \put(-10.6,234.2){.} \put(-10.6,235.2){.} \put(-10.6,236.2){.} \put(-10.6,237.2){.} \put(-10.6,238.2){.} \put(-10.6,239.2){.} \put(-10.5,240.2){.} \put(-10.5,241.2){.} \put(-10.5,242.2){.} \put(-10.5,243.2){.} \put(-10.5,244.2){.} \put(-10.5,245.2){.} \put(-10.5,246.2){.} \put(-10.5,247.2){.} \put(-10.5,248.2){.} \put(-10.4,249.2){.} \put(-10.4,250.2){.} \put(-10.4,251.2){.} \put(-10.4,252.2){.} \put(-10.4,253.3){.} \put(-10.4,254.3){.} \put(-10.4,255.3){.} \put(-10.4,256.3){.} \put(-10.3,257.3){.} \put(-10.3,258.3){.} \put(-10.3,259.3){.} \put(-10.3,260.3){.} \put(-10.3,261.3){.} \put(-10.3,262.3){.} \put(-10.3,263.3){.} \put(-10.3,264.3){.} \put(-10.3,265.3){.} \put(-10.3,266.3){.} \put(-10.2,267.3){.} \put(-10.2,268.3){.} \put(-10.2,269.3){.} \put(-10.2,270.3){.} \put(-10.2,271.3){.} \put(-10.2,272.3){.} \put(-10.2,273.3){.} \put(-10.2,274.3){.} \put(-10.2,275.3){.} \put(-10.2,276.3){.} \put(-10.1,277.4){.} \put(-10.1,278.4){.} \put(-10.1,279.4){.} \put(-10.1,280.4){.} \put(-10.1,281.4){.} \put(-10.1,282.4){.} \put(-10.1,283.4){.} \put(-10.1,284.4){.} \put(-10.1,285.4){.} \put(-10.1,286.4){.} \put(-10.1,287.4){.} \put(-10.0,288.4){.} \put(-10.0,289.4){.} \put(-10.0,290.4){.} \put(-10.0,291.4){.} \put(-10.0,292.4){.} \put(-10.0,293.4){.} \put(-10.0,294.4){.} \put(-10.0,295.4){.} \put(-10.0,296.4){.} \put(-10.0,297.4){.} \put(-10.0,298.4){.} \put(-10.0,299.4){.}

\multiput(-5.0,-3)(0,0.5){6}{.} %max %fourpi2N=39860.474089 %A2=-0.002490 %A1=0.049905

\put(-14,-10){\scriptsize{$-1$}}

\put(-2.3,299.0){.} \put(-2.3,298.0){.} \put(-2.3,297.0){.} \put(-2.3,296.0){.} \put(-2.3,295.0){.} \put(-2.3,294.0){.} \put(-2.3,293.0){.} \put(-2.3,292.0){.} \put(-2.2,291.0){.} \put(-2.2,290.0){.} \put(-2.2,289.0){.} \put(-2.2,288.0){.} \put(-2.2,287.0){.} \put(-2.2,286.0){.} \put(-2.2,285.0){.} \put(-2.2,284.0){.} \put(-2.2,283.1){.} \put(-2.2,282.1){.} \put(-2.2,281.1){.} \put(-2.2,280.1){.} \put(-2.2,279.1){.} \put(-2.2,278.1){.} \put(-2.2,277.1){.} \put(-2.2,276.1){.} \put(-2.2,275.1){.} \put(-2.2,274.1){.} \put(-2.2,273.1){.} \put(-2.2,272.1){.} \put(-2.2,271.1){.} \put(-2.2,270.1){.} \put(-2.2,269.1){.} \put(-2.2,268.1){.} \put(-2.2,267.1){.} \put(-2.2,266.1){.} \put(-2.2,265.1){.} \put(-2.2,264.1){.} \put(-2.2,263.1){.} \put(-2.2,262.1){.} \put(-2.2,261.1){.} \put(-2.2,260.1){.} \put(-2.2,259.1){.} \put(-2.2,258.1){.} \put(-2.2,257.1){.} \put(-2.2,256.1){.} \put(-2.2,255.1){.} \put(-2.2,254.1){.} \put(-2.2,253.1){.} \put(-2.2,252.1){.} \put(-2.2,251.1){.} \put(-2.2,250.1){.} \put(-2.2,249.1){.} \put(-2.2,248.1){.} \put(-2.2,247.1){.} \put(-2.1,246.1){.} \put(-2.1,245.1){.} \put(-2.1,244.1){.} \put(-2.1,243.1){.} \put(-2.1,242.1){.} \put(-2.1,241.1){.} \put(-2.1,240.1){.} \put(-2.1,239.1){.} \put(-2.1,238.2){.} \put(-2.1,237.2){.} \put(-2.1,236.2){.} \put(-2.1,235.2){.} \put(-2.1,234.2){.} \put(-2.1,233.2){.} \put(-2.1,232.2){.} \put(-2.1,231.2){.} \put(-2.1,230.2){.} \put(-2.1,229.2){.} \put(-2.1,228.2){.} \put(-2.1,227.2){.} \put(-2.1,226.2){.} \put(-2.1,225.2){.} \put(-2.1,224.2){.} \put(-2.1,223.2){.} \put(-2.1,222.2){.} \put(-2.1,221.2){.} \put(-2.1,220.2){.} \put(-2.1,219.2){.} \put(-2.1,218.2){.} \put(-2.1,217.2){.} \put(-2.1,216.2){.} \put(-2.1,215.2){.} \put(-2.1,214.2){.} \put(-2.1,213.2){.} \put(-2.1,212.2){.} \put(-2.1,211.2){.} \put(-2.1,210.2){.} \put(-2.0,209.2){.} \put(-2.0,208.2){.} \put(-2.0,207.2){.} \put(-2.0,206.2){.} \put(-2.0,205.2){.} \put(-2.0,204.2){.} \put(-2.0,203.2){.} \put(-2.0,202.2){.} \put(-2.0,201.2){.} \put(-2.0,200.3){.} \put(-2.0,199.3){.} \put(-2.0,198.3){.} \put(-2.0,197.3){.} \put(-2.0,196.3){.} \put(-2.0,195.3){.} \put(-2.0,194.3){.} \put(-2.0,193.3){.} \put(-2.0,192.3){.} \put(-2.0,191.3){.} \put(-2.0,190.3){.} \put(-2.0,189.3){.} \put(-2.0,188.3){.} \put(-2.0,187.3){.} \put(-2.0,186.3){.} \put(-2.0,185.3){.} \put(-2.0,184.3){.} \put(-2.0,183.3){.} \put(-2.0,182.3){.} \put(-2.0,181.3){.} \put(-2.0,180.3){.} \put(-2.0,179.3){.} \put(-1.9,178.3){.} \put(-1.9,177.3){.} \put(-1.9,176.3){.} \put(-1.9,175.3){.} \put(-1.9,174.3){.} \put(-1.9,173.3){.} \put(-1.9,172.3){.} \put(-1.9,171.3){.} \put(-1.9,170.3){.} \put(-1.9,169.4){.} \put(-1.9,168.4){.} \put(-1.9,167.4){.} \put(-1.9,166.4){.} \put(-1.9,165.4){.} \put(-1.9,164.4){.} \put(-1.9,163.4){.} \put(-1.9,162.4){.} \put(-1.9,161.4){.} \put(-1.9,160.4){.} \put(-1.9,159.4){.} \put(-1.9,158.4){.} \put(-1.9,157.4){.} \put(-1.9,156.4){.} \put(-1.9,155.4){.} \put(-1.9,154.4){.} \put(-1.9,153.4){.} \put(-1.8,152.4){.} \put(-1.8,151.4){.} \put(-1.8,150.4){.} \put(-1.8,149.4){.} \put(-1.8,148.4){.} \put(-1.8,147.4){.} \put(-1.8,146.4){.} \put(-1.8,145.4){.} \put(-1.8,144.4){.} \put(-1.8,143.4){.} \put(-1.8,142.5){.} \put(-1.8,141.5){.} \put(-1.8,140.5){.} \put(-1.8,139.5){.} \put(-1.8,138.5){.} \put(-1.8,137.5){.} \put(-1.8,136.5){.} \put(-1.8,135.5){.} \put(-1.8,134.5){.} \put(-1.8,133.5){.} \put(-1.8,132.5){.} \put(-1.7,131.5){.} \put(-1.7,130.5){.} \put(-1.7,129.5){.} \put(-1.7,128.5){.} \put(-1.7,127.5){.} \put(-1.7,126.5){.} \put(-1.7,125.5){.} \put(-1.7,124.5){.} \put(-1.7,123.5){.} \put(-1.7,122.5){.} \put(-1.7,121.5){.} \put(-1.7,120.5){.} \put(-1.7,119.6){.} \put(-1.7,118.6){.} \put(-1.7,117.6){.} \put(-1.7,116.6){.} \put(-1.7,115.6){.} \put(-1.7,114.6){.} \put(-1.6,113.6){.} \put(-1.6,112.6){.} \put(-1.6,111.6){.} \put(-1.6,110.6){.} \put(-1.6,109.6){.} \put(-1.6,108.6){.} \put(-1.6,107.6){.} \put(-1.6,106.6){.} \put(-1.6,105.6){.} \put(-1.6,104.6){.} \put(-1.6,103.6){.} \put(-1.6,102.6){.} \put(-1.6,101.6){.} \put(-1.6,100.7){.} \put(-1.6,99.7){.} \put(-1.6,98.7){.} \put(-1.5,97.7){.} \put(-1.5,96.7){.} \put(-1.5,95.7){.} \put(-1.5,94.7){.} \put(-1.5,93.7){.} \put(-1.5,92.7){.} \put(-1.5,91.7){.} \put(-1.5,90.7){.} \put(-1.5,89.7){.} \put(-1.5,88.7){.} \put(-1.5,87.7){.} \put(-1.5,86.7){.} \put(-1.5,85.7){.} \put(-1.4,84.8){.} \put(-1.4,83.8){.} \put(-1.4,82.8){.} \put(-1.4,81.8){.} \put(-1.4,80.8){.} \put(-1.4,79.8){.} \put(-1.4,78.8){.} \put(-1.4,77.8){.} \put(-1.4,76.8){.} \put(-1.4,75.8){.} \put(-1.4,74.8){.} \put(-1.3,73.8){.} \put(-1.3,72.8){.} \put(-1.3,71.9){.} \put(-1.3,70.9){.} \put(-1.3,69.9){.} \put(-1.3,68.9){.} \put(-1.3,67.9){.} \put(-1.3,66.9){.} \put(-1.3,65.9){.} \put(-1.2,64.9){.} \put(-1.2,63.9){.} \put(-1.2,62.9){.} \put(-1.2,61.9){.} \put(-1.2,60.9){.} \put(-1.2,60.0){.} \put(-1.2,59.0){.} \put(-1.2,58.0){.} \put(-1.1,57.0){.} \put(-1.1,56.0){.} \put(-1.1,55.0){.} \put(-1.1,54.0){.} \put(-1.1,53.0){.} \put(-1.1,52.0){.} \put(-1.1,51.0){.} \put(-1.0,50.1){.} \put(-1.0,49.1){.} \put(-1.0,48.1){.} \put(-1.0,47.1){.} \put(-1.0,46.1){.} \put(-1.0,45.1){.} \put(-0.9,44.1){.} \put(-0.9,43.1){.} \put(-0.9,42.2){.} \put(-0.9,41.2){.} \put(-0.9,40.2){.} \put(-0.8,39.2){.} \put(-0.8,38.2){.} \put(-0.8,37.2){.} \put(-0.8,36.2){.} \put(-0.8,35.3){.} \put(-0.7,34.3){.} \put(-0.7,33.3){.} \put(-0.7,32.3){.} \put(-0.7,31.3){.} \put(-0.6,30.4){.} \put(-0.6,29.4){.} \put(-0.6,28.4){.} \put(-0.5,27.4){.} \put(-0.5,26.4){.} \put(-0.5,25.5){.} \put(-0.4,24.5){.} \put(-0.4,23.5){.} \put(-0.4,22.5){.} \put(-0.3,21.6){.} \put(-0.3,20.6){.} \put(-0.2,19.6){.} \put(-0.2,18.6){.} \put(-0.2,17.7){.} \put(-0.1,16.7){.} \put(-0.0,15.7){.} \put(0.0,14.8){.} \put(0.1,13.8){.} \put(0.1,12.9){.} \put(0.2,11.9){.} \put(0.3,10.9){.} \put(0.4,10.0){.} \put(0.5,9.1){.} \put(0.6,8.1){.} \put(0.7,7.2){.} \put(0.9,6.3){.} \put(1.0,5.4){.} \put(1.2,4.5){.} \put(1.5,3.6){.} \put(1.8,2.8){.} \put(2.2,2.0){.} \put(2.7,1.4){.} \put(3.4,0.8){.} \put(4.3,0.4){.} \put(5.2,0.2){.} \put(6.2,0.1){.} \put(7.2,0.1){.} \put(8.2,0.0){.} \put(9.2,0.0){.} \put(10.2,0.0){.} \put(11.2,0.0){.} \put(12.2,0.0){.} \put(13.2,0.0){.} \put(14.2,0.0){.} \put(15.2,0.0){.} \put(16.2,0.0){.} \put(17.2,0.0){.} \put(18.2,0.0){.} \put(19.2,0.0){.} \put(20.2,0.0){.} \put(21.2,0.0){.} \put(22.2,0.0){.} \put(23.2,0.0){.} \put(24.2,0.0){.} \put(25.2,0.0){.} \put(26.2,0.0){.} \put(27.2,0.0){.} \put(28.2,0.0){.} \put(29.2,0.0){.} \put(30.2,0.0){.} \put(31.2,0.0){.} \put(32.2,0.0){.} \put(33.2,0.0){.} \put(34.2,0.0){.} \put(35.2,0.0){.} \put(36.2,0.0){.} \put(37.2,0.0){.} \put(38.2,0.0){.} \put(39.2,0.0){.}

%frameHeight=300.000000 %A20=-2.000000 %frameStep=1.000000 %gamma=10.000000 %A2max=20.000000 %horizontalFrameToA2Scale=8.000000 %verticalFrameTo4pi2Nscale=15.000000 %horizontalFrameShift=70.000000

\put(54.0,0.0){.} \put(55.0,0.0){.} \put(56.0,0.0){.} \put(57.0,0.0){.} \put(58.0,0.1){.} \put(59.0,0.1){.} \put(60.0,0.1){.} \put(61.0,0.2){.} \put(62.0,0.3){.} \put(63.0,0.4){.} \put(64.0,0.6){.} \put(64.9,1.0){.} \put(65.9,1.5){.} \put(66.7,2.2){.} \put(67.3,3.1){.} \put(67.8,4.1){.} \put(68.2,5.2){.} \put(68.6,6.2){.} \put(68.8,7.3){.} \put(69.1,8.3){.} \put(69.3,9.4){.} \put(69.4,10.4){.} \put(69.6,11.5){.} \put(69.7,12.5){.} \put(69.8,13.6){.} \put(70.0,14.6){.} \put(70.1,15.6){.} \put(70.2,16.7){.} \put(70.2,17.7){.} \put(70.3,18.7){.} \put(70.4,19.8){.} \put(70.5,20.8){.} \put(70.6,21.8){.} \put(70.6,22.8){.} \put(70.7,23.9){.} \put(70.7,24.9){.} \put(70.8,25.9){.} \put(70.9,26.9){.} \put(70.9,27.9){.} \put(71.0,29.0){.} \put(71.0,30.0){.} \put(71.1,31.0){.} \put(71.1,32.0){.} \put(71.1,33.0){.} \put(71.2,34.0){.} \put(71.2,35.1){.} \put(71.3,36.1){.} \put(71.3,37.1){.} \put(71.3,38.1){.} \put(71.4,39.1){.} \put(71.4,40.1){.} \put(71.4,41.2){.} \put(71.5,42.2){.} \put(71.5,43.2){.} \put(71.5,44.2){.} \put(71.6,45.2){.} \put(71.6,46.2){.} \put(71.6,47.2){.} \put(71.6,48.2){.} \put(71.7,49.3){.} \put(71.7,50.3){.} \put(71.7,51.3){.} \put(71.7,52.3){.} \put(71.8,53.3){.} \put(71.8,54.3){.} \put(71.8,55.3){.} \put(71.8,56.3){.} \put(71.9,57.3){.} \put(71.9,58.4){.} \put(71.9,59.4){.} \put(71.9,60.4){.} \put(71.9,61.4){.} \put(72.0,62.4){.} \put(72.0,63.4){.} \put(72.0,64.4){.} \put(72.0,65.4){.} \put(72.0,66.4){.} \put(72.1,67.4){.} \put(72.1,68.4){.} \put(72.1,69.4){.} \put(72.1,70.5){.} \put(72.1,71.5){.} \put(72.1,72.5){.} \put(72.2,73.5){.} \put(72.2,74.5){.} \put(72.2,75.5){.} \put(72.2,76.5){.} \put(72.2,77.5){.} \put(72.2,78.5){.} \put(72.3,79.5){.} \put(72.3,80.5){.} \put(72.3,81.5){.} \put(72.3,82.5){.} \put(72.3,83.6){.} \put(72.3,84.6){.} \put(72.3,85.6){.} \put(72.4,86.6){.} \put(72.4,87.6){.} \put(72.4,88.6){.} \put(72.4,89.6){.} \put(72.4,90.6){.} \put(72.4,91.6){.} \put(72.4,92.6){.} \put(72.5,93.6){.} \put(72.5,94.6){.} \put(72.5,95.6){.} \put(72.5,96.6){.} \put(72.5,97.6){.} \put(72.5,98.6){.} \put(72.5,99.7){.} \put(72.5,100.7){.} \put(72.5,101.7){.} \put(72.6,102.7){.} \put(72.6,103.7){.} \put(72.6,104.7){.} \put(72.6,105.7){.} \put(72.6,106.7){.} \put(72.6,107.7){.} \put(72.6,108.7){.} \put(72.6,109.7){.} \put(72.6,110.7){.} \put(72.7,111.7){.} \put(72.7,112.7){.} \put(72.7,113.7){.} \put(72.7,114.7){.} \put(72.7,115.7){.} \put(72.7,116.7){.} \put(72.7,117.8){.} \put(72.7,118.8){.} \put(72.7,119.8){.} \put(72.7,120.8){.} \put(72.8,121.8){.} \put(72.8,122.8){.} \put(72.8,123.8){.} \put(72.8,124.8){.} \put(72.8,125.8){.} \put(72.8,126.8){.} \put(72.8,127.8){.} \put(72.8,128.8){.} \put(72.8,129.8){.} \put(72.8,130.8){.} \put(72.8,131.8){.} \put(72.9,132.8){.} \put(72.9,133.8){.} \put(72.9,134.8){.} \put(72.9,135.8){.} \put(72.9,136.8){.} \put(72.9,137.8){.} \put(72.9,138.8){.} \put(72.9,139.8){.} \put(72.9,140.9){.} \put(72.9,141.9){.} \put(72.9,142.9){.} \put(72.9,143.9){.} \put(72.9,144.9){.} \put(73.0,145.9){.} \put(73.0,146.9){.} \put(73.0,147.9){.} \put(73.0,148.9){.} \put(73.0,149.9){.} \put(73.0,150.9){.} \put(73.0,151.9){.} \put(73.0,152.9){.} \put(73.0,153.9){.} \put(73.0,154.9){.} \put(73.0,155.9){.} \put(73.0,156.9){.} \put(73.0,157.9){.} \put(73.0,158.9){.} \put(73.1,159.9){.} \put(73.1,160.9){.} \put(73.1,161.9){.} \put(73.1,162.9){.} \put(73.1,163.9){.} \put(73.1,164.9){.} \put(73.1,165.9){.} \put(73.1,167.0){.} \put(73.1,168.0){.} \put(73.1,169.0){.} \put(73.1,170.0){.} \put(73.1,171.0){.} \put(73.1,172.0){.} \put(73.1,173.0){.} \put(73.1,174.0){.} \put(73.1,175.0){.} \put(73.2,176.0){.} \put(73.2,177.0){.} \put(73.2,178.0){.} \put(73.2,179.0){.} \put(73.2,180.0){.} \put(73.2,181.0){.} \put(73.2,182.0){.} \put(73.2,183.0){.} \put(73.2,184.0){.} \put(73.2,185.0){.} \put(73.2,186.0){.} \put(73.2,187.0){.} \put(73.2,188.0){.} \put(73.2,189.0){.} \put(73.2,190.0){.} \put(73.2,191.0){.} \put(73.2,192.0){.} \put(73.3,193.0){.} \put(73.3,194.0){.} \put(73.3,195.0){.} \put(73.3,196.0){.} \put(73.3,197.0){.} \put(73.3,198.0){.} \put(73.3,199.1){.} \put(73.3,200.1){.} \put(73.3,201.1){.} \put(73.3,202.1){.} \put(73.3,203.1){.} \put(73.3,204.1){.} \put(73.3,205.1){.} \put(73.3,206.1){.} \put(73.3,207.1){.} \put(73.3,208.1){.} \put(73.3,209.1){.} \put(73.3,210.1){.} \put(73.3,211.1){.} \put(73.3,212.1){.} \put(73.4,213.1){.} \put(73.4,214.1){.} \put(73.4,215.1){.} \put(73.4,216.1){.} \put(73.4,217.1){.} \put(73.4,218.1){.} \put(73.4,219.1){.} \put(73.4,220.1){.} \put(73.4,221.1){.} \put(73.4,222.1){.} \put(73.4,223.1){.} \put(73.4,224.1){.} \put(73.4,225.1){.} \put(73.4,226.1){.} \put(73.4,227.1){.} \put(73.4,228.1){.} \put(73.4,229.1){.} \put(73.4,230.1){.} \put(73.4,231.1){.} \put(73.4,232.1){.} \put(73.4,233.1){.} \put(73.4,234.1){.} \put(73.5,235.1){.} \put(73.5,236.2){.} \put(73.5,237.2){.} \put(73.5,238.2){.} \put(73.5,239.2){.} \put(73.5,240.2){.} \put(73.5,241.2){.} \put(73.5,242.2){.} \put(73.5,243.2){.} \put(73.5,244.2){.} \put(73.5,245.2){.} \put(73.5,246.2){.} \put(73.5,247.2){.} \put(73.5,248.2){.} \put(73.5,249.2){.} \put(73.5,250.2){.} \put(73.5,251.2){.} \put(73.5,252.2){.} \put(73.5,253.2){.} \put(73.5,254.2){.} \put(73.5,255.2){.} \put(73.5,256.2){.} \put(73.5,257.2){.} \put(73.5,258.2){.} \put(73.6,259.2){.} \put(73.6,260.2){.} \put(73.6,261.2){.} \put(73.6,262.2){.} \put(73.6,263.2){.} \put(73.6,264.2){.} \put(73.6,265.2){.} \put(73.6,266.2){.} \put(73.6,267.2){.} \put(73.6,268.2){.} \put(73.6,269.2){.} \put(73.6,270.2){.} \put(73.6,271.2){.} \put(73.6,272.2){.} \put(73.6,273.2){.} \put(73.6,274.2){.} \put(73.6,275.2){.} \put(73.6,276.2){.} \put(73.6,277.2){.} \put(73.6,278.2){.} \put(73.6,279.2){.} \put(73.6,280.2){.} \put(73.6,281.3){.} \put(73.6,282.3){.} \put(73.6,283.3){.} \put(73.6,284.3){.} \put(73.6,285.3){.} \put(73.7,286.3){.} \put(73.7,287.3){.} \put(73.7,288.3){.} \put(73.7,289.3){.} \put(73.7,290.3){.} \put(73.7,291.3){.} \put(73.7,292.3){.} \put(73.7,293.3){.} \put(73.7,294.3){.} \put(73.7,295.3){.} \put(73.7,296.3){.} \put(73.7,297.3){.} \put(73.7,298.3){.} \put(73.7,299.3){.}

\multiput(77.9,-3)(0,0.5){6}{.} %max %fourpi2N=2466.580581 %A2=0.984682 %A1=0.992312

\put(77.9,-10){\scriptsize{$+1$}}

\put(85.1,299.2){.} \put(85.1,298.2){.} \put(85.1,297.2){.} \put(85.1,296.2){.} \put(85.2,295.2){.} \put(85.2,294.2){.} \put(85.2,293.2){.} \put(85.2,292.2){.} \put(85.2,291.2){.} \put(85.2,290.2){.} \put(85.2,289.2){.} \put(85.2,288.2){.} \put(85.2,287.2){.} \put(85.3,286.2){.} \put(85.3,285.2){.} \put(85.3,284.3){.} \put(85.3,283.3){.} \put(85.3,282.3){.} \put(85.3,281.3){.} \put(85.3,280.3){.} \put(85.3,279.3){.} \put(85.4,278.3){.} \put(85.4,277.3){.} \put(85.4,276.3){.} \put(85.4,275.3){.} \put(85.4,274.3){.} \put(85.4,273.3){.} \put(85.4,272.3){.} \put(85.4,271.3){.} \put(85.5,270.3){.} \put(85.5,269.3){.} \put(85.5,268.3){.} \put(85.5,267.3){.} \put(85.5,266.3){.} \put(85.5,265.3){.} \put(85.5,264.3){.} \put(85.5,263.3){.} \put(85.6,262.3){.} \put(85.6,261.3){.} \put(85.6,260.3){.} \put(85.6,259.3){.} \put(85.6,258.3){.} \put(85.6,257.3){.} \put(85.6,256.3){.} \put(85.7,255.4){.} \put(85.7,254.4){.} \put(85.7,253.4){.} \put(85.7,252.4){.} \put(85.7,251.4){.} \put(85.7,250.4){.} \put(85.8,249.4){.} \put(85.8,248.4){.} \put(85.8,247.4){.} \put(85.8,246.4){.} \put(85.8,245.4){.} \put(85.8,244.4){.} \put(85.8,243.4){.} \put(85.9,242.4){.} \put(85.9,241.4){.} \put(85.9,240.4){.} \put(85.9,239.4){.} \put(85.9,238.4){.} \put(85.9,237.4){.} \put(86.0,236.4){.} \put(86.0,235.4){.} \put(86.0,234.4){.} \put(86.0,233.4){.} \put(86.0,232.4){.} \put(86.0,231.4){.} \put(86.1,230.5){.} \put(86.1,229.5){.} \put(86.1,228.5){.} \put(86.1,227.5){.} \put(86.1,226.5){.} \put(86.2,225.5){.} \put(86.2,224.5){.} \put(86.2,223.5){.} \put(86.2,222.5){.} \put(86.2,221.5){.} \put(86.3,220.5){.} \put(86.3,219.5){.} \put(86.3,218.5){.} \put(86.3,217.5){.} \put(86.3,216.5){.} \put(86.4,215.5){.} \put(86.4,214.5){.} \put(86.4,213.5){.} \put(86.4,212.5){.} \put(86.4,211.5){.} \put(86.5,210.5){.} \put(86.5,209.6){.} \put(86.5,208.6){.} \put(86.5,207.6){.} \put(86.5,206.6){.} \put(86.6,205.6){.} \put(86.6,204.6){.} \put(86.6,203.6){.} \put(86.6,202.6){.} \put(86.7,201.6){.} \put(86.7,200.6){.} \put(86.7,199.6){.} \put(86.7,198.6){.} \put(86.8,197.6){.} \put(86.8,196.6){.} \put(86.8,195.6){.} \put(86.8,194.6){.} \put(86.9,193.6){.} \put(86.9,192.7){.} \put(86.9,191.7){.} \put(86.9,190.7){.} \put(87.0,189.7){.} \put(87.0,188.7){.} \put(87.0,187.7){.} \put(87.0,186.7){.} \put(87.1,185.7){.} \put(87.1,184.7){.} \put(87.1,183.7){.} \put(87.2,182.7){.} \put(87.2,181.7){.} \put(87.2,180.7){.} \put(87.2,179.7){.} \put(87.3,178.7){.} \put(87.3,177.8){.} \put(87.3,176.8){.} \put(87.4,175.8){.} \put(87.4,174.8){.} \put(87.4,173.8){.} \put(87.5,172.8){.} \put(87.5,171.8){.} \put(87.5,170.8){.} \put(87.6,169.8){.} \put(87.6,168.8){.} \put(87.6,167.8){.} \put(87.7,166.8){.} \put(87.7,165.8){.} \put(87.8,164.9){.} \put(87.8,163.9){.} \put(87.8,162.9){.} \put(87.9,161.9){.} \put(87.9,160.9){.} \put(88.0,159.9){.} \put(88.0,158.9){.} \put(88.0,157.9){.} \put(88.1,156.9){.} \put(88.1,155.9){.} \put(88.2,155.0){.} \put(88.2,154.0){.} \put(88.3,153.0){.} \put(88.3,152.0){.} \put(88.4,151.0){.} \put(88.4,150.0){.} \put(88.5,149.0){.} \put(88.5,148.0){.} \put(88.6,147.1){.} \put(88.6,146.1){.} \put(88.7,145.1){.} \put(88.7,144.1){.} \put(88.8,143.1){.} \put(88.9,142.1){.} \put(88.9,141.1){.} \put(89.0,140.2){.} \put(89.1,139.2){.} \put(89.1,138.2){.} \put(89.2,137.2){.} \put(89.3,136.2){.} \put(89.3,135.3){.} \put(89.4,134.3){.} \put(89.5,133.3){.} \put(89.6,132.3){.} \put(89.7,131.3){.} \put(89.8,130.4){.} \put(89.9,129.4){.} \put(90.0,128.4){.} \put(90.1,127.5){.} \put(90.2,126.5){.} \put(90.3,125.5){.} \put(90.4,124.6){.} \put(90.5,123.6){.} \put(90.7,122.7){.} \put(90.8,121.7){.} \put(91.0,120.8){.} \put(91.2,119.9){.} \put(91.4,119.0){.} \put(91.7,118.2){.} \put(92.1,117.5){.} \put(92.7,117.2){.}

%\multiput(92.7,-3)(0,0.5){6}{.} %min %fourpi2N=7.811037 %A2=2.835593 %A1=1.683922

\put(93.6,118.7){.} \put(94.0,119.7){.} \put(94.2,120.8){.} \put(94.5,121.8){.} \put(94.7,122.8){.} \put(94.9,123.9){.} \put(95.1,124.9){.} \put(95.2,125.9){.} \put(95.4,126.9){.} \put(95.5,127.9){.} \put(95.7,128.9){.} \put(95.8,129.9){.} \put(96.0,130.9){.} \put(96.1,131.9){.} \put(96.2,132.9){.} \put(96.4,133.9){.} \put(96.5,134.9){.} \put(96.6,135.9){.} \put(96.7,136.9){.} \put(96.9,137.9){.} \put(97.0,138.9){.} \put(97.1,139.9){.} \put(97.2,140.9){.} \put(97.4,141.9){.} \put(97.5,142.9){.} \put(97.6,143.9){.} \put(97.7,144.8){.} \put(97.9,145.8){.} \put(98.0,146.8){.} \put(98.1,147.8){.} \put(98.3,148.7){.} \put(98.4,149.7){.} \put(98.6,150.6){.} \put(98.8,151.6){.} \put(98.9,152.5){.} \put(99.1,153.4){.} \put(99.4,154.2){.} \put(99.8,154.8){.}

\multiput(99.8,-3)(0,0.5){6}{.} %max %fourpi2N=10.319524 %A2=3.722105 %A1=1.929276

\put(99.8,-10){\scriptsize{$+4$}}

\put(100.6,153.2){.} \put(100.8,152.1){.} \put(101.0,151.0){.} \put(101.2,150.0){.} \put(101.3,148.9){.} \put(101.4,147.9){.} \put(101.5,146.9){.} \put(101.6,145.8){.} \put(101.7,144.8){.} \put(101.8,143.8){.} \put(101.9,142.8){.} \put(102.0,141.8){.} \put(102.1,140.8){.} \put(102.1,139.7){.} \put(102.2,138.7){.} \put(102.3,137.7){.} \put(102.4,136.7){.} \put(102.4,135.7){.} \put(102.5,134.7){.} \put(102.6,133.7){.} \put(102.7,132.7){.} \put(102.7,131.7){.} \put(102.8,130.7){.} \put(102.9,129.7){.} \put(102.9,128.7){.} \put(103.0,127.7){.} \put(103.0,126.7){.} \put(103.1,125.7){.} \put(103.2,124.7){.} \put(103.2,123.7){.} \put(103.3,122.7){.} \put(103.4,121.7){.} \put(103.4,120.7){.} \put(103.5,119.7){.} \put(103.6,118.7){.} \put(103.6,117.7){.} \put(103.7,116.7){.} \put(103.7,115.7){.} \put(103.8,114.7){.} \put(103.9,113.7){.} \put(103.9,112.7){.} \put(104.0,111.7){.} \put(104.0,110.7){.} \put(104.1,109.7){.} \put(104.2,108.7){.} \put(104.2,107.7){.} \put(104.3,106.7){.} \put(104.3,105.7){.} \put(104.4,104.7){.} \put(104.5,103.7){.} \put(104.5,102.7){.} \put(104.6,101.7){.} \put(104.7,100.7){.} \put(104.7,99.7){.} \put(104.8,98.7){.} \put(104.8,97.7){.} \put(104.9,96.7){.} \put(105.0,95.7){.} \put(105.0,94.7){.} \put(105.1,93.8){.} \put(105.2,92.8){.} \put(105.2,91.8){.} \put(105.3,90.8){.} \put(105.4,89.8){.} \put(105.4,88.8){.} \put(105.5,87.8){.} \put(105.6,86.8){.} \put(105.6,85.8){.} \put(105.7,84.8){.} \put(105.8,83.8){.} \put(105.9,82.8){.} \put(105.9,81.8){.} \put(106.0,80.8){.} \put(106.1,79.9){.} \put(106.2,78.9){.} \put(106.2,77.9){.} \put(106.3,76.9){.} \put(106.4,75.9){.} \put(106.5,74.9){.} \put(106.5,73.9){.} \put(106.6,72.9){.} \put(106.7,71.9){.} \put(106.8,70.9){.} \put(106.9,70.0){.} \put(107.0,69.0){.} \put(107.1,68.0){.} \put(107.1,67.0){.} \put(107.2,66.0){.} \put(107.3,65.0){.} \put(107.4,64.0){.} \put(107.5,63.0){.} \put(107.6,62.1){.} \put(107.7,61.1){.} \put(107.8,60.1){.} \put(107.9,59.1){.} \put(108.0,58.1){.} \put(108.1,57.1){.} \put(108.3,56.2){.} \put(108.4,55.2){.} \put(108.5,54.2){.} \put(108.6,53.2){.} \put(108.7,52.2){.} \put(108.8,51.3){.} \put(109.0,50.3){.} \put(109.1,49.3){.} \put(109.2,48.3){.} \put(109.4,47.4){.} \put(109.5,46.4){.} \put(109.7,45.4){.} \put(109.8,44.4){.} \put(110.0,43.5){.} \put(110.2,42.5){.} \put(110.3,41.5){.} \put(110.5,40.6){.} \put(110.7,39.6){.} \put(110.9,38.6){.} \put(111.1,37.7){.} \put(111.3,36.7){.} \put(111.5,35.8){.} \put(111.7,34.8){.} \put(112.0,33.9){.} \put(112.2,32.9){.} \put(112.5,32.0){.} \put(112.8,31.1){.} \put(113.0,30.1){.} \put(113.3,29.2){.} \put(113.7,28.3){.} \put(114.0,27.4){.} \put(114.4,26.5){.} \put(114.8,25.6){.} \put(115.2,24.7){.} \put(115.6,23.9){.} \put(116.1,23.0){.} \put(116.6,22.2){.} \put(117.2,21.5){.} \put(117.8,20.7){.} \put(118.5,20.0){.} \put(119.2,19.3){.} \put(119.9,18.8){.} \put(120.8,18.2){.} \put(121.6,17.8){.} \put(122.5,17.4){.} \put(123.5,17.1){.} \put(124.4,17.0){.} \put(125.4,16.9){.}

%\multiput(125.4,-3)(0,0.5){6}{.} %min %fourpi2N=1.126120 %A2=6.929853 %A1=2.632461

\put(126.4,16.9){.} \put(127.4,17.0){.} \put(128.4,17.1){.} \put(129.4,17.3){.} \put(130.4,17.5){.} \put(131.4,17.8){.} \put(132.3,18.0){.} \put(133.3,18.3){.} \put(134.3,18.5){.} \put(135.2,18.7){.} \put(136.2,18.8){.} \put(137.2,18.8){.}

\multiput(137.2,-3)(0,0.5){6}{.} %max %fourpi2N=1.255324 %A2=8.403773 %A1=2.898926

\put(137.2,-10){\scriptsize{$+9$}}

\put(138.2,18.7){.} \put(139.2,18.5){.} \put(140.2,18.2){.} \put(141.1,17.8){.} \put(142.0,17.3){.} \put(142.9,16.7){.} \put(143.7,16.1){.} \put(144.5,15.5){.} \put(145.3,14.9){.} \put(146.1,14.3){.} \put(146.8,13.7){.} \put(147.6,13.0){.} \put(148.4,12.4){.} \put(149.2,11.8){.} \put(150.0,11.3){.} \put(150.8,10.7){.} \put(151.7,10.2){.} \put(152.5,9.6){.} \put(153.4,9.2){.} \put(154.3,8.7){.} \put(155.2,8.3){.} \put(156.1,7.9){.} \put(157.0,7.5){.} \put(157.9,7.2){.} \put(158.9,6.9){.} \put(159.8,6.6){.} \put(160.8,6.3){.} \put(161.8,6.1){.} \put(162.7,5.9){.} \put(163.7,5.7){.} \put(164.7,5.5){.} \put(165.7,5.4){.} \put(166.7,5.2){.} \put(167.7,5.1){.} \put(168.7,5.0){.} \put(169.6,4.9){.} \put(170.6,4.8){.} \put(171.6,4.8){.} \put(172.6,4.7){.} \put(173.6,4.7){.} \put(174.6,4.6){.} \put(175.6,4.6){.} \put(176.6,4.6){.} \put(177.6,4.6){.} \put(178.6,4.5){.} \put(179.6,4.5){.} \put(180.6,4.5){.} \put(181.6,4.5){.} \put(182.6,4.5){.} \put(183.6,4.5){.} \put(184.6,4.5){.} \put(185.6,4.5){.} \put(186.6,4.5){.} \put(187.6,4.5){.} \put(188.6,4.5){.} \put(189.6,4.4){.} \put(190.6,4.4){.} \put(191.6,4.4){.} \put(192.6,4.3){.} \put(193.6,4.3){.} \put(194.6,4.2){.} \put(195.6,4.2){.} \put(196.6,4.1){.} \put(197.6,4.0){.} \put(198.6,3.9){.} \put(199.6,3.8){.} \put(200.6,3.8){.} \put(201.6,3.7){.} \put(202.6,3.6){.} \put(203.6,3.5){.} \put(204.6,3.4){.} \put(205.6,3.3){.} \put(206.6,3.2){.} \put(207.6,3.1){.} \put(208.6,3.0){.} \put(209.6,2.9){.} \put(210.6,2.8){.} \put(211.6,2.7){.} \put(212.6,2.6){.} \put(213.6,2.6){.} \put(214.6,2.5){.} \put(215.6,2.4){.} \put(216.6,2.4){.} \put(217.5,2.3){.} \put(218.5,2.2){.} \put(219.5,2.2){.}

\end{picture}
\caption{Dependence of $\N$ on $\rv^2$ for small and large values of $\gamma$. }\label{gamma}
\end{figure}
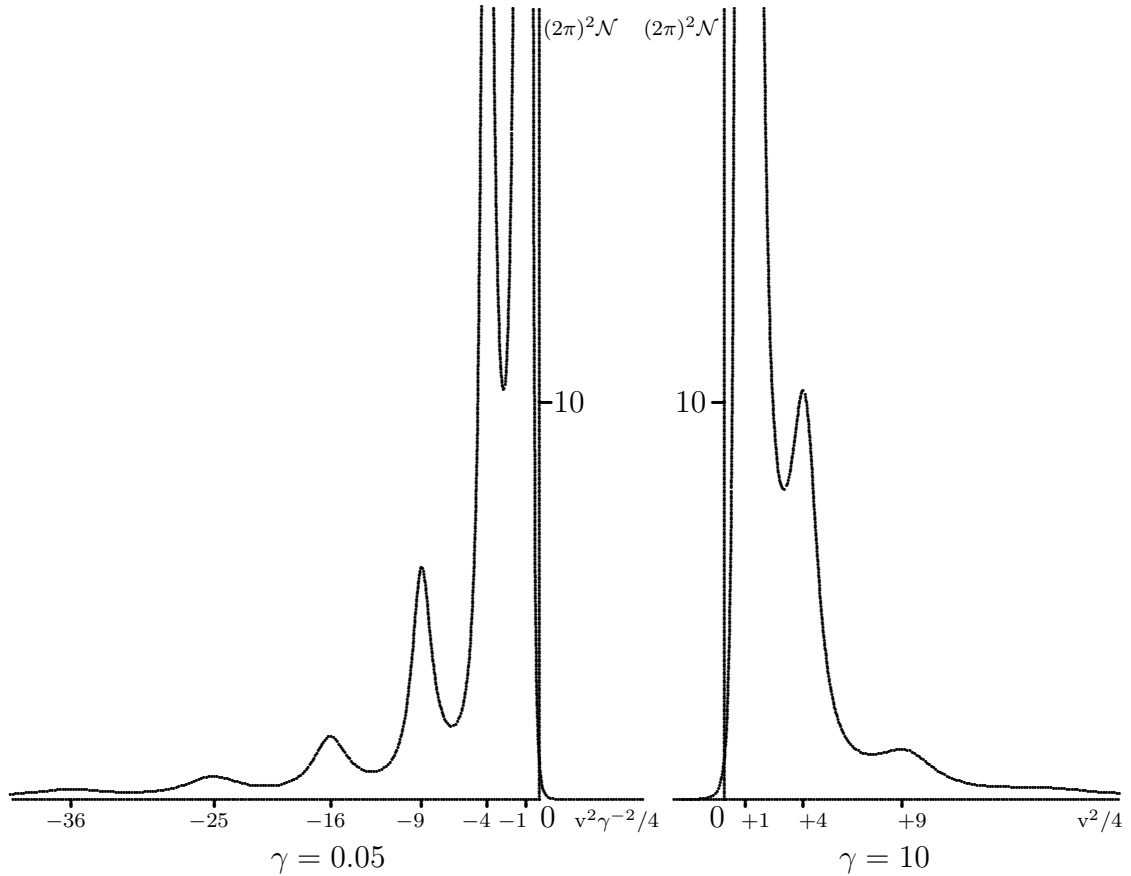

Effect of the considered singular points on behavior of $\N$ in physical region grows especially at $\gamma \ll 1$ or at $\gamma \gg 1$, when these points $\rv^2 = 4 (1 + i / \gamma )^{-2} n^2$, $n = 1, 2, ...$ approach physical region $\Im \rv^2 = 0$. This displays as appearance of a set of local maxima of $\N$ approximately at $\rv^2 = -4 \gamma^2 n^2$, $\gamma \ll 1$ or at $\rv^2 = 4 n^2$, $\gamma \gg 1$ where $n = 1, 2, ...$, see Figure \ref{gamma}.
There $-n^2$ at $\gamma \ll 1$ or $n^2$ at $\gamma \gg 1$, $n = 1, 2, ...$, are approximate values of $\rv^2 \gamma^{-2}/4$ or $\rv^2/4$, respectively, corresponding to the position of local maxima. Taking into account that $\rv$ is triangle area $A$, the maxima are located at $|A| = 2 \gamma n$ or at $|A| = 2 n$ in the spacelike or timelike region, respectively.

As for asymptotic behavior in physical region at $|A| \to \infty$, $\N$ decays as $\exp (-\pi |A|)$ in spacelike region or as $\exp (-\pi |A|/\gamma)$ in timelike region (at $\harcz$; at $\hz$ the exponents are $\exp (- |A|)$ and $\exp (- |A|/\gamma)$, respectively).

\medskip

\noindent {\bf\large 5.Conclusion}

\noindent The above considered moments could have transparent physical sense in a theory with independent area tensors $\vstw$. The theory with independent {\it scalar} areas is known as area Regge calculus \citeup{BarRocWil,RegWil}. Now we can speak of the area {\it tensor} Regge calculus. It is in many respects analogous to the 3 dimensional Regge calculus, and we can find the form of the full discrete path integral \citeup{Kha3} which becomes true canonical one in the formal continuous time limit irrespectively of the coordinate chosen as time just as in the 3 dimensional case mentioned at the end of {\bf Introduction}. Then the vacuum expectation values of the area tensor monomials are just the considered moments (\ref{Malbegade}) at $h(z) = z$ generalized to the integrals of $\N (\{\bvstw\}, \{\bvstw^*\})$ (\ref{intDOm}) with monomials over $\bvstw, \bvstw^*$ for the whole set of 2-simplices $\stw$. Important point in this consideration is that the set of holonomies $\{ \Rstw : \stw \supset \son \}$ for a given link $\son$ obey Bianchi identities \cite{Regge}. Then integration over $\prod_{\stw \supset \son} \d^6 \bvstw$ in the path integral will result in the singularity of the type of $(\desi (\br_{\stw} ))^2$ for some $\stw$. Rather, integration over certain subset of area tensors ${\cal F}$ should be omitted (a kind of gauge fixing). Correspondingly, the moments in general case can be defined as integrals of $\N$ with area tensor monomials over $\d^6 \bvstw$, $\stw \not\in {\cal F}$. At $\harcz$ restrictions on the set of area tensor components to integrate over are relaxed, and this set might be larger than $\{ \bvstw : \stw \not\in {\cal F}\}$; this point requires further studying.

To resume, direct definition of (nonabsolutely convergent) path integral in a theory with finite SO(3,1) rotations should be made with care (mainly because of exponential growth of the Haar measure on Lorentz boosts). In the particular case of such theory, Regge calculus in terms of rotation matrices in Minkowsky spacetime, path integral can be well defined. This definition respects correspondence with Euclidean version. Upon integrating out connections, probability distribution turns out to decay exponentially at large areas. Thus vertices do not go away to infinity and in this sense the minisuperspace system described by elementary lengths/areas is self-consistent.

\medskip

\noindent {\bf\large Acknowledgements}

\noindent The present work was supported in part by the Russian Foundation for Basic Research through Grants No. 08-02-00960-a and No. 09-01-00142-a.

\end{document}